\documentclass[aip,pop,superscriptaddress,reprint]{revtex4-1}

\usepackage[plainpages=false, colorlinks=true, anchorcolor=blue, linkcolor=blue, citecolor=blue, bookmarks=false]{hyperref}

\usepackage{amsmath, amsthm, amssymb}
\usepackage{graphicx}
\usepackage{grfext}

\newcommand{\beq}{\begin{equation}}
\newcommand{\eeq}{\end{equation}}
\newcommand{\bea}{\begin{eqnarray}}
\newcommand{\eea}{\end{eqnarray}}

\newcommand{\bB}{\mathbf{B}}
\newcommand{\bA}{\mathbf{A}}
\newcommand{\ba}{\mathbf{a}}
\newcommand{\bb}{\mathbf{b}}
\newcommand{\bn}{\mathbf{n}}
\newcommand{\bp}{\mathbf{p}}

\newcommand{\bk}{\mathbf{k}}
\newcommand{\bq}{\mathbf{q}}

\newcommand{\bw}{\mathbf{w}}
\newcommand{\bx}{\mathbf{x}}
\newcommand{\by}{\mathbf{y}}
\newcommand{\bz}{\mathbf{z}}
\newcommand{\br}{\mathbf{r}}
\newcommand{\btimes}{\pmb{\times}}

\begin{document}

\title{Inferring Morphology and Strength of Magnetic Fields From Proton Radiographs}

\author{Carlo Graziani}
\email{carlo@oddjob.uchicago.edu}
\affiliation{Flash Center for Computational Science, University of Chicago,
5640 S. Ellis Avenue, Chicago, IL 60637, USA}

\author{Petros Tzeferacos}
\affiliation{Flash Center for Computational Science, University of Chicago,
5640 S. Ellis Avenue, Chicago, IL 60637, USA}

\author{Donald Q. Lamb}
\affiliation{Flash Center for Computational Science, University of Chicago,
5640 S. Ellis Avenue, Chicago, IL 60637, USA}

\author{Chikang Li}
\affiliation{Plasma Science and Fusion Center, Massachusetts Institute 
of Technology, NW17-239, 175 Albany St., Cambridge, MA 02139, USA}

\begin{abstract}
Proton radiography is an important diagnostic method for laser plasma 
experiments, and is particularly important in the analysis of magnetized 
plasmas.  The theory of radiographic image analysis has heretofore only 
permitted somewhat limited analysis of the radiographs of such plasmas.  We 
furnish here a theory that remedies this deficiency. We show that to linear 
order in magnetic field gradients, proton radiographs are projection images of 
the MHD current along the proton trajectories.  We demonstrate that in the 
linear approximation, the full structure of the perpedicular magnetic field can 
be reconstructed by solving a steady-state inhomogeneous 2-dimensional 
diffusion equation sourced by the radiograph fluence contrast data.  We explore 
limitations of the inversion method due to Poisson noise, to discretization 
errors, to radiograph edge effects, and to obstruction by laser target 
structures.  We also provide a separate analysis that is well-suited to the 
inference of isotropic-homogeneous magnetic turbulence spectra.  We 
discuss extension of these results to the nonlinear contrast regime.
\end{abstract}
\date{28 March 2016}
\maketitle

\section{Introduction}

Proton radiography is an experimental technique wherein the electric 
and magnetic fields in a plasma are imaged using beams of protons. 
The deflection by electric and magnetic fields of a proton's path 
bears information about the morphology and strengths 
of electric and magnetic fields. This kind of imaging is in use at 
laser facilities around the world, and has advanced our 
understanding of flows, instabilities, and electric and magnetic 
fields in high energy density plasmas.

Two distinct types of proton sources have been developed for use in 
proton imaging. The first type of source uses thermally-produced 
protons accelerated by the extreme electric field gradients 
generated when a target is illuminated by an intense laser 
\cite{Borghesi2001,Mackinnon2006,Snavely2000,Borghesi2002}. The second 
type of source uses mono-energetic, fusion-produced  protons created 
by the implosion of a D$^3$He capsule \cite{Li2006,Li2000,Seguin2002,Seguin2004}.  

The ability of proton imaging to provide information about the 
strength and morphology of electric and magnetic fields is 
exceptional. Such fields are ubiquitous in high energy density 
physics (HEDP) experiments, since laser-driven flows create strong 
electric fields and the mis-aligned electron density and temperature 
gradients in such flows create strong magnetic fields via the 
Biermann battery effect. Thermally produced protons have been used, 
for example, to detect electric fields in laser-plasma interaction 
experiments \cite{Borghesi2001}; and to image solitons 
\cite{borghesi2002macroscopic}, laser-driven implosions \cite{Mackinnon2006}, and 
collisionless electrostatic shocks \cite{Romagnani2008}. 
Mono-energetic protons have been used, for example to distinguish 
between, image, and quantify electric and magnetic fields in 
laser-driven plasmas from foils \cite{Li2006} and inside hohlraums 
\cite{Li2009}; to study changes in magnetic field topology due to 
reconnection \cite{Li2007}, the morphology of the magnetic fields 
generated in laser-driven implosions \cite{Rygg2008,Li2008,Li2010}, 
the evolution of filamentation and self-generated fields in the 
coronae of directly-driven inertial-confinement-fusion capsules 
\cite{Seguin2012}, and Rayleigh-Taylor induced magnetic fields 
\cite{Manuel2012}; and to observe magnetic field generation via the 
Weibel instability in collisionless shocks \cite{Huntington2015}.

Curiously, proton imaging was used in laser plasma experiments for about a 
decade before any theory was deveoped regarding how such images should be 
analyzed to recover the electromagnetic structure of the imaged plasma.  Prior 
to the publication of the paper by Kugland et al.\cite{kugland2012a} (hereafter 
K2012), essentially no quantitative information could be extracted from proton 
radiographs.  The K2012 paper furnished the first systematic quantitative 
discussion of the analysis of proton radiographs, identifying the principal 
physical parameters, and identifying the ``linear'' (small image contrast) and 
``nonlinear'' (large image contrast) regimes.  K2012 noted the frequent 
occurrence of high-definition structures in proton images, and highlighted 
caustics of the proton optical system in the nonlinear regime as the key to 
interpreting such structures. Starting from a selection of simplified field 
configurations, K2012 explored the caustic image structures to be expected in 
radiographs of those configurations, and advocated for an approach that relies 
on recognizing the typology of such structures, and using them to infer the 
fields that produce them.

In this paper, we provide a first-principles description of the nature of the 
images of magnetized plasmas produced by proton imaging, and discuss the 
implications for gaining information about the morphology and the strength of 
magnetic fields in high energy density laboratory experiments.  Our concern is 
strictly with the imaging of magnetic fields; we do not treat the imaging of 
electric fields.  

We mainly treat linear contrast fluctuations in this work, although 
we also give a schematic discussion of how the nonlinear regime can 
be addressed. In the linear regime, as we show below, image 
contrasts in proton radiographs are \textit{images of projected MHD 
current.} It follows from this observation that it is perfectly 
possible for high-definition structure to occur even in the linear 
regime. This is the case because where currents with sharp 
distributions occur, sharp features must occur in their radiographic 
images, irrespective of whether focusing is strong or weak. We 
follow up the brief discussion in K2012 that showed that in the 
linear regime, inversion possible in principle, by furnishing and 
verifying a practical field reconstruction algorithm, and 
separately, an algorithm for recovering turbulent spectra.

In \S\ref{sec:PR_DandM} we describe the experimental setup, 
including the proton imaging diagnostic, that we consider in this 
paper.  In \S\ref{sec:SFT} we provide a theory of proton radiography 
in the limit of small image contrasts, a situation that applies in 
many HEDP experiments. In \S\ref{sec:reconstruction} we show how the 
proton radiographic image can be used to reconstruct the average 
transverse magnetic field using this theory. We present verification 
tests of the theory in \S\ref{sec:verification}.  In \S\ref{sec:errors}, 
we explore and characterize some of the errors that 
affect the reconstruction of the magnetic field using proton 
radiographs.  In \S\ref{sec:turbulence} we apply the theory to 
reconstruction of the spectrum of the turbulent magnetic field 
produced by a turbulent flow.  In \S\ref{sec:Nonlinear}, we outline 
schematically an approach that we believe may permit extension of 
the field reconstruction algorithm to the nonlinear regime.  We 
discuss our results and draw some conclusions in \S\ref{sec:Discussion}.

\section{Proton Radiography: Definitions and Magnitudes}\label{sec:PR_DandM}

\begin{figure*}[t]
\begin{center}
\includegraphics[scale=0.6, viewport=0 0 637 233]{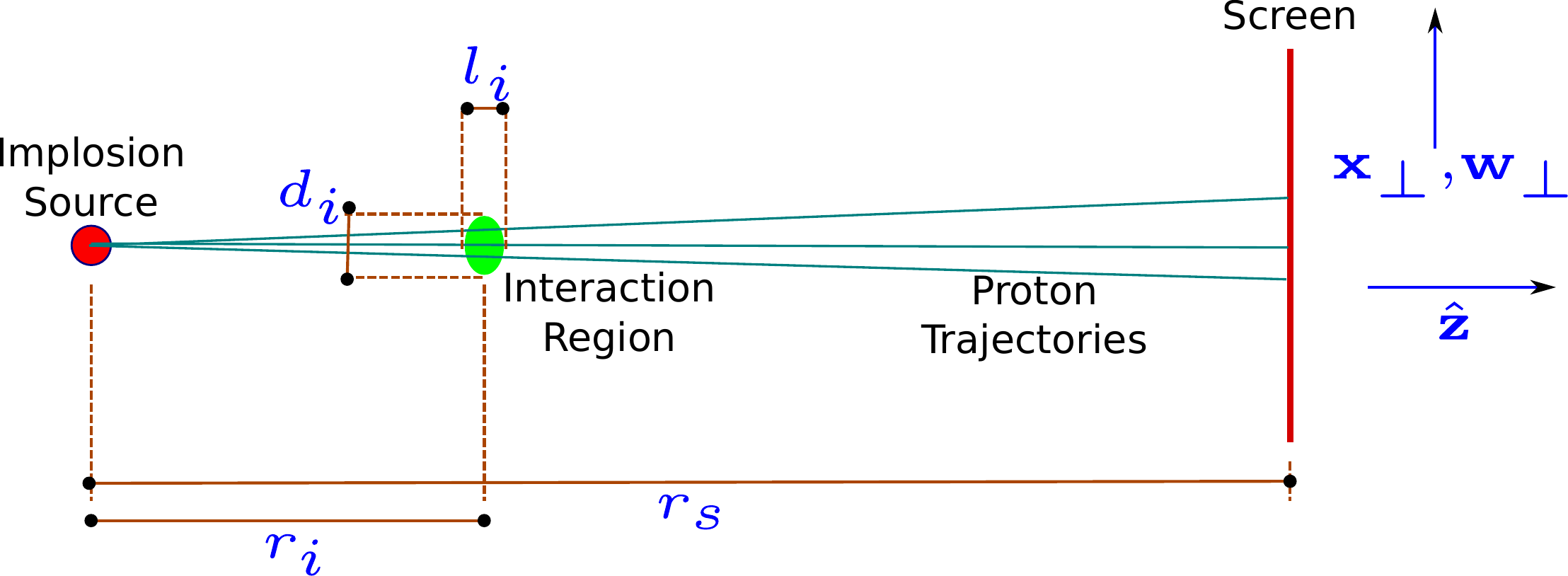}
\end{center}
\caption{Illustration of Experimental Setup}
\label{Fig:SetupIllustration}
\end{figure*}

In this work, we concern ourselves exclusively with proton-radiographic imaging 
of magnetic structures, neglecting all electric forces on proton beams, unlike
the more general investigation of K2012\cite{kugland2012a}.

An illustration of the experimental setup for a proton radiographic image 
system is shown in Figure~\ref{Fig:SetupIllustration}. Protons emitted 
isotropically from the implosion source at the left of the figure pass through 
an interaction region of non-zero magnetic field, where they suffer small 
deflections due to the Lorentz force. They then continue on,  striking a 
screen, where their positions are recorded. 

If the transverse size of the interaction region $d_{i}$ is assumed small 
compared to the distance $r_{i}$ of the interaction region from the implosion 
source, then the proton trajectories are nearly parallel. That is, if we assume 
that the angular spread $\omega\equiv d_{i}/r_{i}\ll 1$ -- the ``paraxial'' 
assumption also made in K2012 -- we may use planar geometry at 
the screen.  We will adopt this assumption in what follows.

We further assume that the longitudinal size of the interaction region, $l_i$, 
is small compared to $r_i$, so that the parameter $\lambda\equiv l_i/r_i\ll 1$. 
This assumption is convenient because it allows certain geometric factors 
arising in integrals to be treated as constants.  This assumption was 
considered equivalent to the previous assumption $\omega\ll 1$ in 
K2012, since in that work the longitudinal and transverse extents 
of the interaction region were assumed equal.  Since there are two separate 
uses made here of the ``small interaction region'' approximation, one 
(paraxiality) related to the transverse extent, the other related to the 
longitudinal extent, we prefer to distinguish between the two extensions, even
if they are usually physically similar.

A third simplifying assumption made here in common with K2012
is that
the angular deflections $\alpha$ induced by the magnetic field
are very small. We may quantify this conservatively by assuming a
uniform field strength $B$, which gives rise to a Larmor radius $r_{B}=\frac{mvc}{eB}$
for protons of mass $m$ moving at velocity $v$. When $r_{B}\gg l_{i}$
we may estimate $\alpha\approx\frac{l_{i}}{r_{B}}=\frac{eBl_{i}}{\beta mc^{2}}$,
where $\beta\equiv\frac{v}{c}$. If the protons have kinetic energy
$T$, then $\beta\approx\sqrt{2T/mc^{2}}$. Therefore,
\bea
\alpha&=&\frac{e}{\sqrt{2mc^{2}}}T^{-1/2}Bl_{i}\nonumber \\
&=& 1.80\times10^{-2}\,\textnormal{Rad}\,\times
       \left(\frac{T}{14.7~\textnormal{MeV}}\right)^{-1/2}\nonumber\\
&& \times\left(\frac{B}{10^5~\textnormal{G}}\right)
   \times\left(\frac{l_{i}}{0.1~\textnormal{cm}}\right).
       \label{eq:Small_Angles}
\eea
With proton kinetic energies of 3~MeV or 14.7~MeV characterstic of D--$^3$He
capsules, and magnetic field
strengths of even $10^6$~Gauss, even a coherent field produces
small angular deflections. A random walk due to a disordered field
would produce even smaller deflections.

K2012 note the importance of a further ``small'' experimental 
parameter, which measures the strength of the variations of angular 
deflection induced by varying fields across the interaction region. 
Assuming a perturbation of length scale $a$ in the transverse 
direction inducing a deflection angle of characteristic size $\alpha$, 
the definition of the dimensionless number $\mu$ is
\beq
\mu\equiv\frac{l_i\alpha}{a}.
\label{eq:mu}
\eeq

In effect, $\mu$ measures the amount of deflection per unit (transverse) length
in the interaction region.  It is clearly closely related to the derivatives of 
the magnetic field.  The particular importance of $\mu$ is that it controls the 
\textit{amplitude} of density contrast on the radiographic image -- small 
$\mu$ corresponds to small-amplitude contrasts, whereas large $\mu$ results 
in order-unity or larger (that is, ``non-linear'') contrasts. It is noted in 
K2012 that $\mu$ and $\alpha$ are independent parameters, which 
may be large or small without reference to each other.  In particular, one may 
have fields that are intense (large $\alpha$) but relatively uniform (small 
$\mu$) yielding small contrasts, or conversely weak fields (small $\alpha$)
with large gradients (large $\mu$) producing large contrasts.  The latter
regime was of interest in K2012, since it is the regime in which caustic
surfaces may occur.

To summarize the research reported in this paper with respect to the 
parameters $\omega$, $\lambda$, $\alpha$, and $\mu$, we assume  that 
$\alpha\ll 1$, $\omega\ll 1$, $\lambda\ll 1$, and $\mu\ll 1$, which 
is commonly the case in many HEDP-related proton-radiographic 
experimental setups. We will see below that in this ``linear'' (in 
$\mu$) regime, $\mu$ is in fact directly related to MHD current 
density $\nabla\btimes \bB$, so that in effect contrast amplitude is 
directly related (in fact, proportional) to current strength. 

As we will see, the identification of $\mu$ with MHD current is 
sufficient to reconstruct the average transverse magnetic field in 
detail, and, separately, to infer the spectra of 
isotropic-homogeneous magnetic turbulent fields.  It also serves as 
a caution with respect to the use of caustic structure in the 
interpretation of radiographs of magnetized plasmas: high-definition 
structure on a radiograph cannot, in general, be attributed to
caustic surfaces of the proton optical system, since they can be 
produced by sharp \textit{current} distributions in regimes where 
magnetic field gradients are too weak to produce caustics. Incautious 
identification of high-definition radiographic ``blobs'' with 
caustics can therefore lead to overestimation of magnetic field 
strengths.

\section{Small Fluctuation Theory of Proton Radiography}\label{sec:SFT}

We now consider the small contrast regime, $\mu\ll 1$.  As we will now 
see, the theory of such contrasts can be constructed in terms of a 
2-dimensional deflection vector field in the focal plane.  The divergence of 
this field yields the proton fluence contrast field directly, and may be 
related to derivatives of the magnetic field in the interaction region -- 
specifically to the MHD current.

\subsection{The Lateral Displacement Field}\label{subsec:LateralDeviations}

The fluence distribution of protons at the screen is represented by an area
density function, $\Psi(\bx_{\perp})\,d^{2}\bx_{\perp}$,
where $\bx_{\perp}$ represents position along the screen.
In the absence of a magnetic field, the density function is uniform
by assumption, $\Psi_{0}(\bx_{\perp})=\psi_{0}$, a constant.
The effect of magnetization in the interaction region is to introduce
a field of small lateral displacements $\bw_{\perp}(\bx_{\perp})$
that distort the uniform field $\Psi_{0}$ to produce $\Psi(\bx_{\perp})$.
The $\bw_{\perp}$ are ``small'' in the sense that they correspond
to the small angular deflections discussed above.

We may relate the change in $\Psi$ due to the magnetic field to the
displacement field $\bw_{\perp}$. If we let the undisturbed coordinates
be $\bx^{(0)}_{\perp}$, so that $\bx_{\perp}=\bx^{(0)}_{\perp}+\bw_{\perp}$,
we have by conservation of protons
\bea
\Psi(\bx_{\perp})\,d^{2}\bx_{\perp}
 & = & \Psi_{0}(\bx^{(0)}_{\perp})d^{2}\bx^{(0)}_{\perp}\nonumber \\
 & = & \Psi_{0}(\bx_{\perp}-\bw_{\perp})\,
    \left|\frac{\partial\bx^{(0)}_{\perp}}{\partial\bx_{\perp}}\right|
    d^{2}\bx_{\perp}\nonumber \\
 & \approx & \left[\Psi_{0}(\bx_{\perp})-\bw_{\perp}\cdot
             \nabla_{\perp}\Psi_{0}(\bx_{\perp})+
             {\cal O}(\alpha^2)\right]\nonumber\\
  &&\times\left[1-\nabla_{\perp}\cdot\bw_{\perp}+{\cal O}(\mu^2)\right]\,
             d^{2}\bx_{\perp},\label{eq:Proton_Cons}
\eea
to first order in $\alpha$ and $\mu$. It follows that 
\beq
\Psi(\bx_{\perp})\approx\Psi_{0}(\bx_{\perp})-\nabla_{\perp}
  \cdot\left[\Psi_{0}(\bx_{\perp})\bw_{\perp}\right],
  \label{eq:Proton_Cons2}
\eeq
which expresses the conservation of the ``flux'' $\Psi_{0}(\bx_{\perp})\bw_{\perp}$
in the plane of the detector screen. Since the unperturbed proton flux
is uniform by assumption, so that $\Psi_{0}(\bx_{\perp})=\psi_{0}$,
a constant, we finally obtain
\bea
\delta\Psi(\bx_{\perp})/\psi_{0} & \equiv &
  \frac{\Psi(\bx_{\perp})-\psi_{0}}{\psi_{0}}\nonumber \\
 & \approx & -\nabla_{\perp}\cdot\bw_{\perp}.
\label{eq:Proton_Cons3}
\eea

Equation (\ref{eq:Proton_Cons3}) specifies our chief observable,
the contrast field 
\beq
\Lambda(\bx_{\perp})\equiv\delta\Psi(\bx_{\perp})/\psi_{0},
\label{eq:Lambda_def}
\eeq
and relates it to the model quantity $\bw_{\perp}$.

It should be clear from the above that we are making essential use
of the conditions $\alpha\ll 1$ and $\mu\ll 1$.
In particular, $|\bw_{\perp}|/r_{s}\sim{\cal O}(\alpha)$, and
$\Lambda\sim{\cal O}(\mu)$, and we
drop all terms of higher order in $\alpha$ and $\mu$.

\subsection{The Lorentz Force}\label{subsec:LorentzForce}

We may relate the lateral displacement field $\bw_{\perp}$ to
the magnetic field in the interaction region. Between its emission
and a later time $t$, a proton experiences a lateral motion $\delta\bx_{\perp}(t)$
given by
\beq
\delta\bx_{\perp}(t)=\int_{0}^{t}dt_{1}\,\mathbf{v}_{\perp}(t_{1}).
\label{eq:Lateral_Motion1}
\eeq
Here, $\mathbf{v}_{\perp}(t_1)$ is the lateral velocity experienced
at time $t_1$ due to the Lorentz force accelerations received at prior
times. It is given by
\beq
\mathbf{v}_{\perp}(t_1)=\int_{0}^{t_1}dt_{2}\,\frac{e}{mc}v\,\bn\btimes
\bB\left(\bx(t_{2})\right),
\label{eq:Lateral_Velocity1}
\eeq
where $\bn$ is the tangent vector to the proton trajectory,
which is approximately constant along the trajectory, due to the smallness
of the angular deflection $\alpha$. 

We treat the implosion source as approximately point-like, and use
the near-constancy of $\bn$, so that the argument $\bx(t)$
of $\bB$ in the integrand of Equation~(\ref{eq:Lateral_Velocity1})
is given by $\bx(t)=\bn vt$. The Lorentz force
is perpendicular to $\bn$, so the longitudinal velocity $v$
is constant, and $t=z/v$, where $z$ denotes distance along the $\bn$-direction.
We change variables from $t$ to $z$, obtaining
\bea
\delta\bx_{\perp}(z) & = & \frac{1}{v}\int_{0}^{z}dz_{1}\,
\mathbf{v}_{\perp}(z_{1})
\label{eq:Lateral_Motion2}\\
\mathbf{v}_{\perp}(z_{1}) & = & \frac{e}{mc}\int_{0}^{z_{1}}dz_{2}\,
\bn\btimes\bB(\bn z_{2}).
\label{eq:Lateral_Velocity2}
\eea

Combining Equations~(\ref{eq:Lateral_Motion2}) and (\ref{eq:Lateral_Velocity2}),
we get
\beq
\delta\bx_{\perp}(z)=\frac{e}{mcv}\bn\btimes
\int_{0}^{z}dz_{1}\int_{0}^{z_{1}}dz_{2}\,\bB(\bn z_{2}).
\label{eq:Lateral_Motion3}
\eeq
We may switch the order of integration, to obtain
\bea
\delta\bx_{\perp}(z) & = & \frac{e}{mcv}\bn\btimes
\int_{0}^{z}dz_{2}\,\bB(\bn z_{2})\int_{z_{2}}^{z}dz_{1}\nonumber \\
 & = & \frac{e}{mcv}\bn\btimes\int_{0}^{z}dz_{2}\,\bB
 (\bn z_{2})\,(z-z_{2}).
 \label{eq:Lateral_Motion4}
\eea
The intuitive meaning of this equation is that the lateral displacement 
$\delta\bx_\perp(z)$ of a proton at $z$ is a superposition of displacements resulting from 
additive velocity pulses $\frac{e}{mc}(\bn v)\btimes\bB\,d(z_2/v)$,
each pulse operating for a time $(z-z_2)/v$.

The relation between $\bn$ and the transverse location at
the screen $\bx_{\perp}$ is
\beq
\bn = \left[\hat{\bz}+\bx_{\perp}/r_{s}\right]
+{\cal O}(\omega^2),
\label{eq:n_z}
\eeq
where, again, $\omega=d_i/r_i$ is the small spread angle of protons crossing
the interaction region. Defining
\beq
\br(z,\bx_{\perp})\equiv
(\hat{\bz}+\bx_{\perp}/r_{s})z
\label{eq:x_z}
\eeq
we may express the lateral deflection at the screen, $\bw_{\perp}=\delta\bx_{\perp}(r_{s})$
to leading order in $\omega$ and $\alpha$:
\beq
\bw_{\perp}(\bx_{\perp})=\frac{e}{mcv}\bn\btimes
\int_{0}^{r_{s}}dz\,\bB\left(\br(z,\mathbf{x_{\perp}})\right)
\,(r_{s}-z).
\label{eq:Lateral_Deviation}
\eeq

As a further simplification, we may take advantage of the fact that
in the integrand of Equation~(\ref{eq:Lateral_Deviation}), the variation
of the magnetic field is much more rapid than that of the factor $(r_{s}-z)$,
which may therefore be taken outside of the integral and replaced
with $(r_{s}-r_{i})$ at the cost of a relative error of order 
$\left|z-r_{i}\right|/r_{i}\sim l_{i}/r_{i}=\lambda$ (this is the ``small 
longitudinal extent'' approximation discussed earlier). The result is
\beq
\mathbf{w_{\perp}}(\bx_{\perp})=\frac{e(r_{s}-r_{i})}{mcv}\bn
\btimes\int_{0}^{r_{s}}dz\,\bB
\left(\br(z,\mathbf{x_{\perp}})\right).
\label{eq:Lateral_Deviation2}
\eeq

\subsection{Divergence of Deviation Field}\label{subsec:Divergence}

From Equation~(\ref{eq:Proton_Cons3}), we know that the proton density
field $\delta\Psi(\bx_{\perp})/\psi_{0}$ is connected to the
quantity $\nabla_{\perp}\cdot\bw_{\perp}$, rather that to
$\bw_{\perp}$ directly. To calculate this quantity, it is
convenient to extend $\bw_{\perp}(\bx_{\perp})$ to
a vector field $\mathbf{u}(\bx)$ defined in a 3-D neighborhood
of the detector screen. The definition of $\mathbf{u}$ is
\beq
\mathbf{u}(\bx)\equiv\frac{e(r_{s}-r_{i})}{mcv}
\frac{\bx}{\left|\bx\right|}\btimes
\int_{0}^{r_{s}}dz\,\bB
\left(\frac{\bx}{\left|\bx\right|}z\right).
\label{eq:uvec}
\eeq
When $\bx=\bx_{\perp}+r_{s}\hat{\bz}$, it follows
that $\left|\mathbf{u}(\bx)-\bw_{\perp}(\bx_{\perp})\right|
/\left|\bw(\bx_{\perp})\right|\sim{\cal O}(\omega)$.
We also have $\nabla\cdot\mathbf{u}=\left(\nabla_{\perp}\cdot
\bw_{\perp}\right)\left(1+{\cal O}(\omega)\right)$,
since the component of $\mathbf{u}$ along $\hat{\bz}$ is
${\cal O}(\omega)$. This is convenient because it is easier to take
unconstrained spatial derivatives of the expression in Equation~(\ref{eq:uvec})
than to take ``perpendicular'' derivatives of the expression in
Equation~(\ref{eq:Lateral_Deviation2}).

We therefore have 
\bea
\Lambda(\bx_{\perp}) & = & -\nabla_{\perp}\cdot
\bw_{\perp}(\bx_{\perp})\nonumber \\
 & = & -\nabla\cdot\mathbf{u}(\bx)\nonumber \\
 & = & \frac{e(r_{s}-r_{i})}{mcv}\frac{\bx}{\left|\bx\right|}
 \cdot\nabla\btimes\int_{0}^{r_{s}}dz\,\bB\left(
 \frac{\bx}{\left|\bx\right|}z\right),\nonumber\\
 \label{eq:dw_1}
\eea
where to get the third line, we've used the identity 
$\nabla\cdot(\mathbf{P}\btimes\mathbf{Q})=\mathbf{Q\cdot\nabla\btimes}
\mathbf{P}-\mathbf{P\cdot\nabla\btimes}\mathbf{Q}$
and the fact that $\nabla\btimes\frac{\bx}{\left|\bx\right|}=0$.

We pass over to tensor notation, denoting a vector $\mathbf{q}$ by
its components $q_{i}$, $i=1,2,3$, and using the Einstein convention
that repeated indices imply summation. Letting 
$\br(z)\equiv\frac{\bx}{\left|\bx\right|}z$,
and taking the derivative inside the integral, we find
\bea
\Lambda(\bx_{\perp}) & = & \frac{e(r_{s}-r_{i})}{mcv}
       \frac{x_{i}}{\left|\bx\right|}\epsilon_{ijk}\int_{0}^{r_{s}}dz\,
       \frac{\partial}{\partial x_{j}}B_{k}\left(\br\right)\nonumber \\
 & = & \frac{e(r_{s}-r_{i})}{mcv}\epsilon_{ijk}
       \frac{x_{i}}{\left|\bx\right|}\times\nonumber\\
    && \int_{0}^{r_{s}}dz\,\frac{\partial}{\partial r_{l}}B_{k}
       \left(\mathbf{\br}\right)\,
       \frac{\partial r_{l}}{\partial x_{j}}\nonumber \\
 & = & \frac{e(r_{s}-r_{i})}{mcv}\epsilon_{ijk}
       \frac{x_{i}}{\left|\bx\right|}\times\nonumber\\
    && \int_{0}^{r_{s}}dz\,\frac{\partial}{\partial r_{l}}B_{k}
       \left(\mathbf{\br}\right)\,z
       \left(\frac{\delta_{jl}}{\left|\bx\right|}-
       \frac{x_{j}x_{l}}{\left|\bx\right|^{3}}\right)\nonumber \\
 & = & \frac{e(r_{s}-r_{i})}{mcv}\epsilon_{ijk}
       \frac{x_{i}}{\left|\bx\right|^{2}}
       \int_{0}^{r_{s}}dz\,z
       \frac{\partial}{\partial r_{j}}B_{k}\left(\br\right)\nonumber \\
 & = & \frac{e(r_{s}-r_{i})}{mcv}\frac{1}{\left|\bx\right|}\bn
       \cdot\int_{0}^{r_{s}}dz\,z
       \nabla\btimes\bB\left(\br(z)\right)\nonumber\\
 & \approx & \frac{er_i(r_{s}-r_{i})}{mcvr_s}\bn\cdot
       \int_{0}^{r_{s}}dz\,
       \nabla\btimes\bB\left(\br(z)\right)\nonumber\\
 & \equiv & \chi(\bx_\perp^{(0)}).\nonumber\\
       \label{eq:dw_2}
\eea
Here, in the first line, we've used the totally antisymmetric Levi-Civita 
tensor $\epsilon_{ijk}$ to express the cross product in tensor form; to get 
from the third to the fourth equality, we've used the fact that 
$\epsilon_{ijk}x_{i}x_{j}=0$ to eliminate the second term in the parentheses of 
the integrand; and in the final line we have appealed to the rapid variation of 
$\bB(\br(z))$ in the interaction region to replace the factor $z$ in the 
integrand by $r_{i}\left(1+{\cal O}(\lambda)\right)$, and take it outside the 
integral, while also setting $|\bx|=r_s\left(1+{\cal O}(\omega)\right)$. 
In the final line we have defined the \textit{current projection 
function} $\chi(\bx_\perp^{(0)})$, a dimensionless function of the 
undeflected coordinates $\bx_\perp^{(0)}$.

Equation~(\ref{eq:dw_2}) is amenable to an interesting interpretation.
The MHD current $\mathbf{J}$ is given by $\frac{c}{4\pi}\nabla\btimes\bB$,
so that the contrast map $\Lambda(\bx_\perp)$ is in fact proportional
to the line integral of $\mathbf{J}$ along the flight path of the
protons. That is, \emph{the image on the screen is basically the projection
of the $z$-component of the MHD current field.}  We can now also interpret the
small contrast regime as the regime of small MHD currents.

It follows that in the small-contrast theory, proton radiographs  of MHD 
plasmas are, for all intents and purposes, projective images of the MHD current 
component along the proton beam. This is a very useful observation, as it 
furnishes an attractive alternative to the procedure of stacking blob-like 
substructures in the interpretation of radiographs, in the manner advocated in 
K2012.

\section{Average Transverse Magnetic Field 
Reconstruction}\label{sec:reconstruction}

An interesting and useful further consequence of the theory developed thus far 
is that it is possible to reconstruct the proton deflections $\bw_\perp$ due to 
a magnetized plasma. By Equation~(\ref{eq:Lateral_Deviation2}), this is 
equivalent to reconstructing the $z$-integrated transverse magnetic field. This 
possibility was recognized, but not exploited, in K2012, and 
the treatment in that work failed to recognize an important issue that must be 
addressed in order for such a reconstruction to be practical, as we now 
discuss.

We begin by writing Equation~(\ref{eq:Proton_Cons3}) as follows:
\beq
-\nabla_{0\perp}\cdot\bw(\bx_\perp^{(0)})=\Lambda(\bx_\perp),
\label{eq:Recon_eq_1}
\eeq
where $\bw(\bx_\perp^{(0)})$ is the lateral deflection as a function 
of undeflected lateral coordinate $\bx_\perp^{(0)}$, and where the 
notation $\nabla_{0\perp}$ refers to differentiation with respect to 
the undeflected coordinates $\bx_\perp^{(0)}$.  The function 
$\bw(\bx_\perp^{(0)})$ is written in this way to emphasize that it
is a function of the unperturbed coordinates (of which it is also a 
perturbation).

The point about the coordinate dependence of $\bw$ is worth 
belaboring because it is the origin of a subtlety in Equation~(\ref
{eq:Recon_eq_1}): the source term on the right-hand side is a 
function of the \textit{perturbed} coordinates 
$\bx_\perp=\bx_\perp^{(0)}+\bw(\bx_\perp^{(0)})$.  We therefore have
\beq
-\nabla_{0\perp}\cdot\bw(\bx_\perp^{(0)})=\Lambda(\bx_\perp^{(0)}+\bw(\bx_\perp^{(0)})).
\label{eq:Recon_eq_2}
\eeq
The subtlety worth emphasizing here is that we may not, in general, neglect
the correction to the argument of $\Lambda$ on the RHS of 
Equation~(\ref{eq:Recon_eq_2}), despite the small-deflection approximation
$\alpha\ll 1$.  The reason is that while the angular deflections may be a small 
fraction of a radian, they may still result in a substantial change in 
$\Lambda$, in regions where gradients of $\Lambda$ are large.

Now, as pointed out in Equation~(78) of K2012, the 
deflection field $\bw(\bx_\perp^{(0)})$ is irrotational, as a 
consequence of the solenoidal character of $\bB$.  That is, there 
exists a scalar function $\phi(\bx_\perp^{(0)})$ such that
\beq
\bw(\bx_\perp^{(0)})=-\nabla_{0\perp}\phi(\bx_\perp^{(0)}).
\label{eq:w_irrotational}
\eeq

Combining Equations~(\ref{eq:Recon_eq_2}) and (\ref{eq:w_irrotational}),
we obtain the field reconstruction equation
\beq
\nabla^2\phi(\by)=\Lambda\left(\by-\nabla\phi(\by)\right),
\label{eq:Recon_eq_3}
\eeq
where we have unburdened the expressions of their superscripts and subscripts 
to rein in the burgeoning notational complexity,  by introducing 
$\by\equiv\bx_\perp^{(0)}$, and by stipulating that all gradients and 
Laplacians are two-dimensional operators in the plane of the detector, and
differentiate with respect to $\by$.

Equation~(\ref{eq:Recon_eq_3}) differs from the reconstruction equation of 
K2012 (Equation 20 of K2012, basically the Poisson equation for the projected 
potential, and its extension to the magnetic case given in Equation 79) by the 
argument of the source term on the RHS.  We have found that as a practical 
matter, the correction to the argument \textit{may not be neglected};
doing so results in a very inaccurate algorithm, because $\Lambda$ can change 
substantially on this scale.  The correction may be treated approximately, 
however, as we now demonstrate.

Equation~(\ref{eq:Recon_eq_3}) is a second-order, elliptical, nonlinear  partial 
differential equation for the scalar function $\phi$, which must 
obviously be solved numerically.  It is not straightforwardly 
solvable in its full nonlinear form.  However, we may expand the 
right-hand side to first order in $\nabla\phi$, obtaining
\beq
\nabla^2\phi(\by)=\Lambda(\by)-\nabla\Lambda(\by)\cdot\nabla\phi(\by).
\label{eq:Recon_eq_4}
\eeq
Multiplying Equation (\ref{eq:Recon_eq_4}) through by the integrating factor 
$\exp(\Lambda(\by))$, we obtain
\beq
\nabla\cdot\left(e^{\Lambda(\by)}\nabla\phi(\by)\right)=\Lambda(\by)e^{\Lambda(\by)}.
\label{eq:Recon_eq_diffusion}
\eeq

Equation~(\ref{eq:Recon_eq_diffusion}) is a steady-state diffusion 
equation, with an inhomogeneous diffusion coefficient $\exp(\Lambda)$, 
and a source term $\Lambda\exp(\Lambda)$.  It may be solved by 
standard numerical methods, as we demonstrate in \S\ref{sec:verification}.

Once a solution has been obtained, it is straighforward to recover 
the deflection field $\bw=-\nabla\phi$.  The projection integral of 
the perpendicular magnetic field may then be obtained using 
Equation~(\ref{eq:Lateral_Deviation2}):
\beq
\int_{0}^{r_{s}}dz\,\bB_\perp
\left(\br(z,\by)\right)=
\frac{mcv}{e(r_s-r_i)}\nabla\phi(\by)\btimes\bz,
\label{eq:Recon_B}
\eeq
where we've made our now usual approximation $\bn^{(0)}\approx\bz$ at 
the cost of a small error ${\cal O}(\omega)$.  The vector product with 
$\bz$ in effect rotates the vector $\nabla\phi$ clockwise by 
90$^\circ$.

We
may use Equation~(\ref{eq:Recon_B}) to estimate the \textit{longitudinally 
averaged transverse field},
\beq
\bar{\bB}_\perp\equiv\frac{1}{l_i}\int_{0}^{r_{s}}dz\,\bB_\perp
\left(\br(z,\by)\right),\label{eq:Field_Average}
\eeq
which is obviously interesting information that is well-worth extracting.  
In order to obtain this average, we clearly need some way of estimating the 
longitudinal extent of the interaction region, $l_i$.  Such an estimate may be
available from a separate diagnostic measurement of the plasma, or from a 
simulation.  Alternatively, one may assume a rough isotropy, implying that the 
transverse extent of the interaction region is approximately the same as the 
longitudinal extent.  Under this assumption, one may use the radiographic data 
to estimate the transverse extent, and parlay this into the required estimate 
of $l_i$.

\begin{figure*}[t]
\begin{center}
\includegraphics[width=7.0in, viewport=175 65 1660 820, clip=]{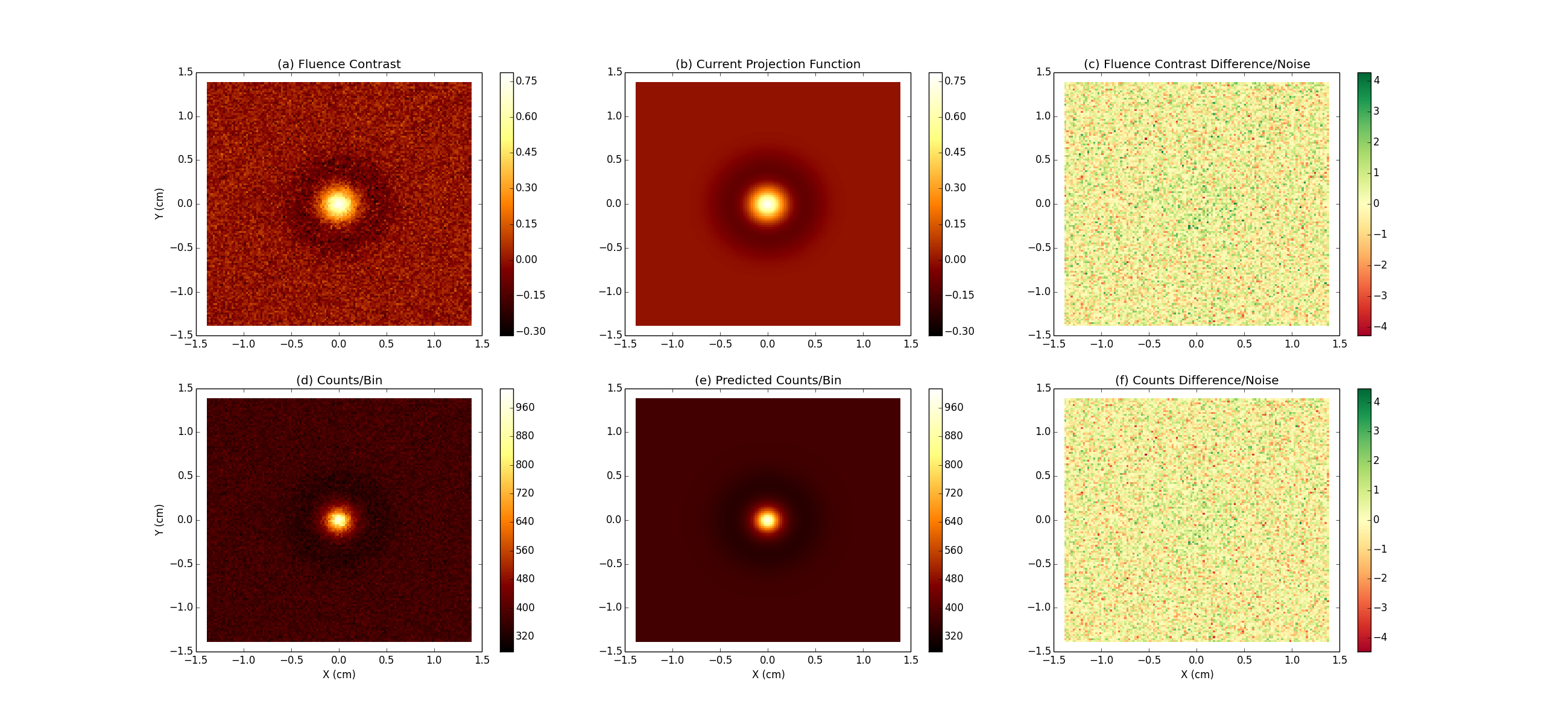}
\caption{\label{figure:base_case_current} Comparison of theory and simulation, 
for an ellipsoidal blob aligned with the $z$-axis. The top row compares the 
fluence contrast to the current projection contrast, testing Equation 
\ref{eq:dw_2}, whereas the bottom row compares the raw counts/bin to the 
predicted counts/bin.  The left panels show the radiograph data, while the 
middle panels show the prediction from the integration of 
Equation~\ref{eq:chi_ode}. The right panels show the result of subtracting the 
middle-panel data from the corresponding right-panel data, and normalizing by 
the statistical error -- the Poisson error $\sqrt{n}$ for the bottom panels, 
the propagated error (Equation~\ref{eq:sigma_L}) for the top panels.  The 
fluctuations in the difference maps are consistent with Gaussian noise.}
\end{center}
\end{figure*}
\section{Numerical Verification of the Theory}\label{sec:verification}

In order to verify the theory developed thus far, we have
written a small ray-tracing code in \textit{Python}, and used it to simulate large 
ensembles of 14.7~MeV protons that are fired at a magnetized region from a 
central source, and recorded at a screen.  The code uses the ODE integration
routines from the \textit{Scipy} \cite{oliphant2007} library to integrate
the trajectory equations
\beq
\frac{d\bn}{ds}=\left(\frac{\alpha}{l_i}\right)\bn\btimes\bb,
\label{eq:dnds_2}
\eeq
and
\beq
\frac{d\bx}{ds}=\bn,
\label{eq:ray_location}
\eeq
as an ODE system. Here, $s$ denotes physical length along each 
integral curve of $\bn$, and we measure the magnetic field in 
units of some typical field strength $B_t$ in the domain, so that 
$\bb\equiv\bB/B_t$, and the deflection angle $\alpha$ is the 
``typical'' deflection $\alpha=\frac{eB_tl_i}{mvc}$.

The ODE system is further augmented by the tracer equations
\beq
\frac{d\chi}{ds}=\frac{er_i(r_s-r_i)}{mcvr_s}\bn\cdot\nabla\btimes\bB\label{eq:chi_ode}
\eeq
and
\beq
\frac{d}{ds}\left(l_i\bar{\bB}_\perp\right)=\bB_\perp,\label{eq:Bperp_ode}
\eeq
which allow us to trace the current projection function $\chi$ and the average
perpendicular field $\bar{\bB}_\perp$ experienced by each proton, so that we can map
them to the screen and compare them with the fluence contrast $\Lambda$
and the reconstructed perpendicular field, respectively, thus verifying the theory.

\begin{figure*}[t]
\begin{center}
\includegraphics[width=5.5in, viewport=95 70 1110 950]{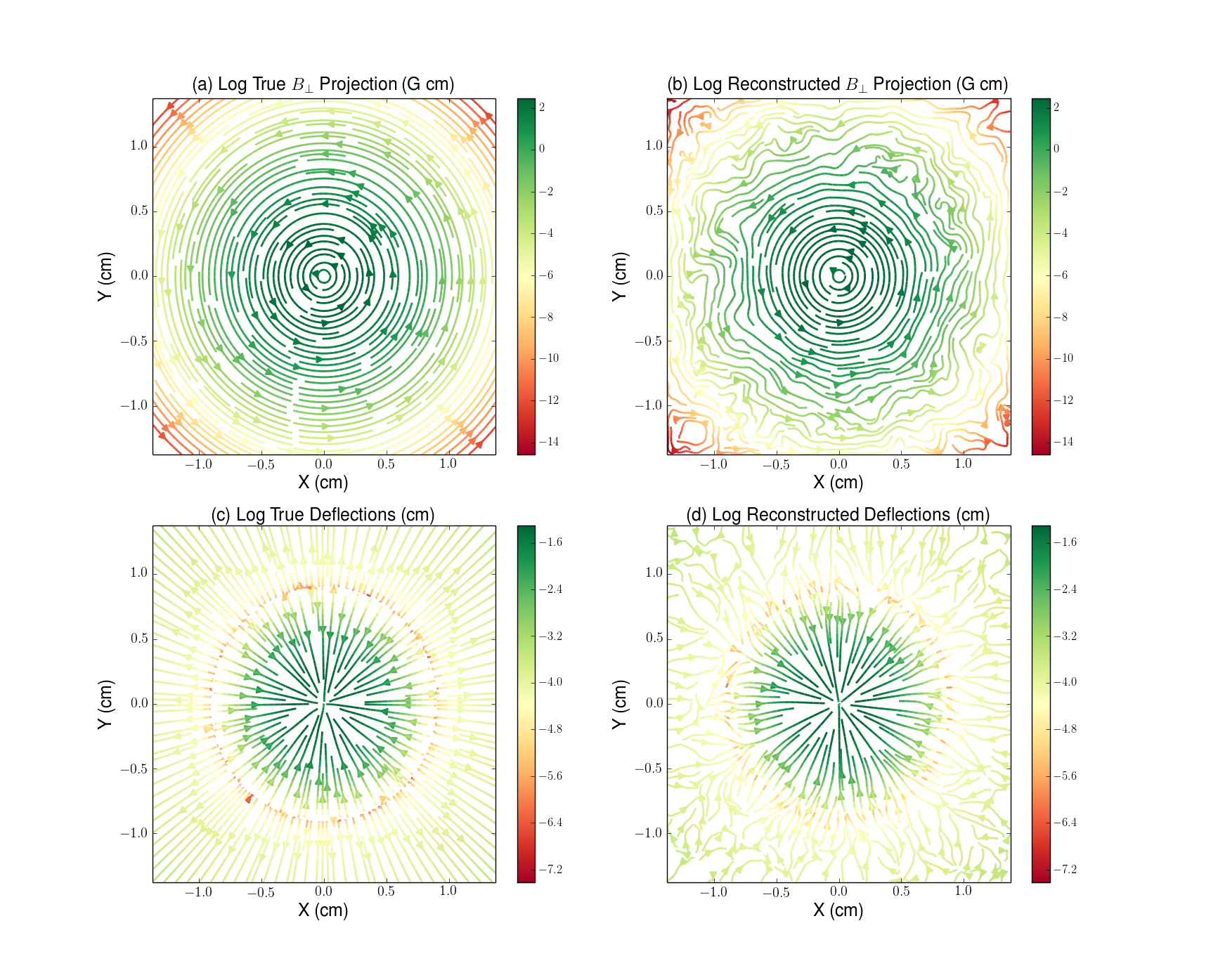}
\caption{\label{figure:base_case_reconstruction}Reconstruction of the field
configuration, for an ellipsoidal blob aligned with the $z$-axis. 
The colorbars displays field strength on a logarithmic scale.
(a) Actual field projection $l_i\bar{\bB}_\perp$ from the integration  of
Equation~(\ref{eq:Bperp_ode}); (b) Reconstructed
field; (c) Actual proton deflections; (d) Reconstructed deflections.
}
\end{center}
\end{figure*}

\begin{figure*}[t]
\begin{center}
\includegraphics[width=7.0in, viewport=175 65 1660 820, clip=]{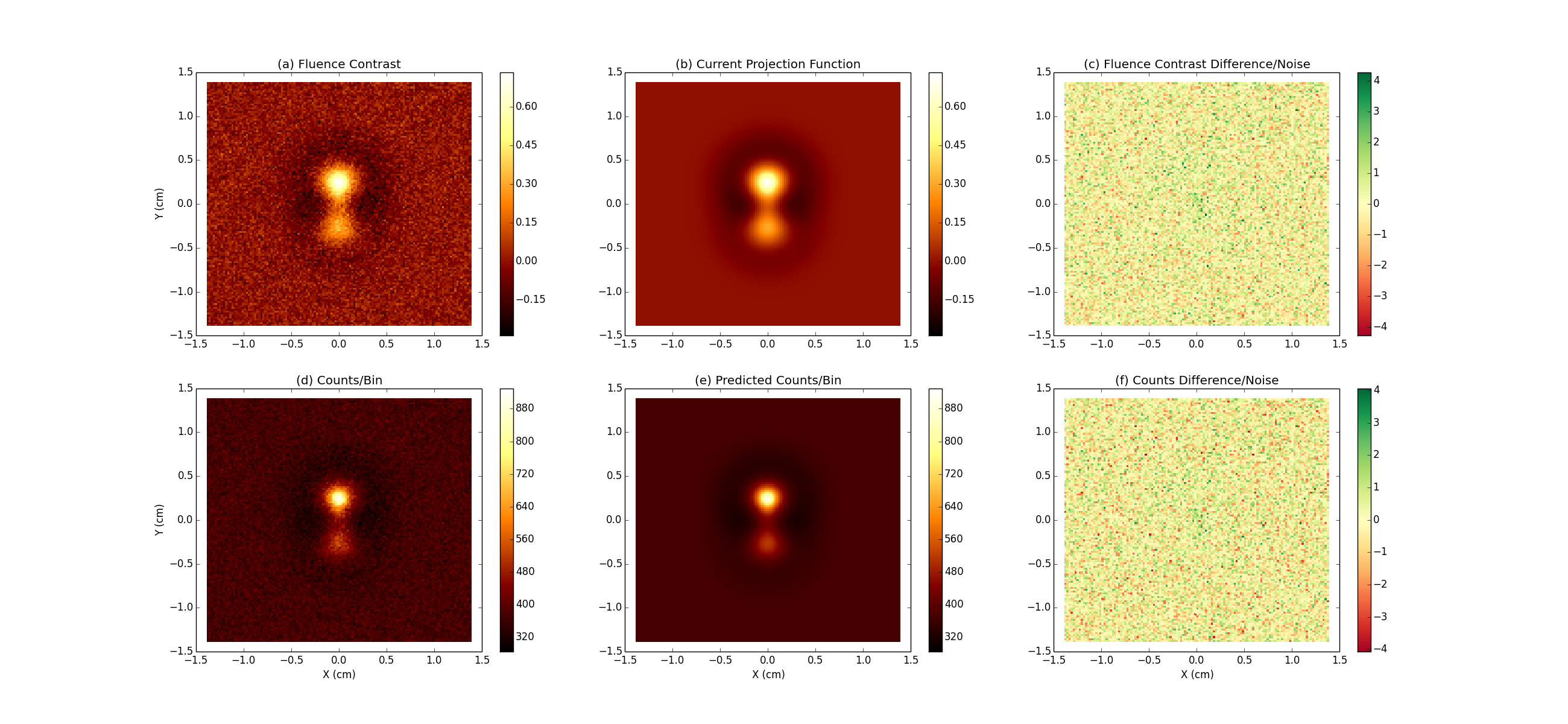}
\caption{\label{figure:twoblob_current}The same as Figure \ref{figure:base_case_current},
but for a superposition of two blobs of unequal field intensity.}
\end{center}
\end{figure*}

\begin{figure*}[t]
\begin{center}
\includegraphics[width=5.5in, viewport=95 70 1110 950]{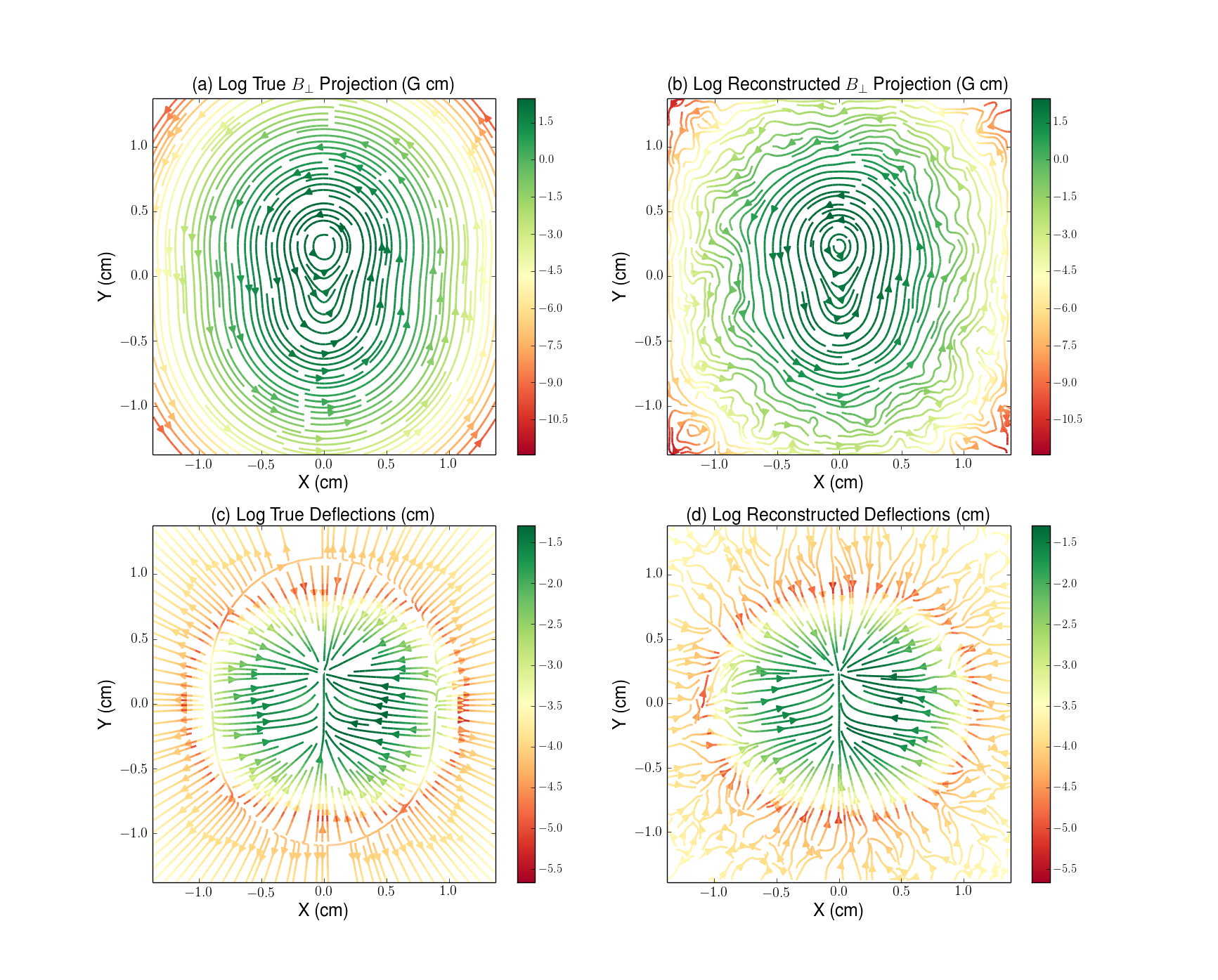}
\caption{\label{figure:twoblob_reconstruction}The same as Figure \ref{figure:base_case_reconstruction},
but for a superposition of two blobs of unequal field intensity.}
\end{center}
\end{figure*}

Protons are fired in a cone about the $z$-axis, bounded by a circular target
in the $x-y$-plane at the location of the interaction region.  The cone is
distorted by the Lorentz force, and the protons locations and tracer data are
recorded at $z=r_s$.  The cone is then ``collimated'' to a square region, with
boundaries aligned with the $x$- and $y$-axes, such that the square is 
entirely contained by the circular aperture of the cone.

We simulate point sources only, located at a distance $r_i=10$~cm from the 
interaction region, and $r_s=100$~cm from the screen.  The magnetic fields that 
we simulate have characteristic field strengths of $2\times 10^4$~G.  The field 
configuration that we simulate is a magnetic ``ellipsoidal blob'', described
in K2012.  The field is purely azimuthal about an arbitrary
axis, with field strength
\beq
B_\phi(r^\prime,z^\prime)=B_0\frac{r^\prime}{a}
\exp\left(-\frac{r^{\prime 2}}{a^2}-\frac{z^{\prime 2}}{b^2}\right),\label{eq:blob}
\eeq
where $r^\prime$ is perpendicular distance from the azimuthal symmetry axis, and
$z^\prime$ is distance along the axis.  The axis may be rotated arbitrarily with
respect to the arrival direction of the protons.

All images are binned into 128$\times$128 pixels.  The protons are summed in 
each bin, while the current projection function $\chi$ and the average 
transverse field $l_i\bar{\bB}_\perp$, integrated for each proton according to 
Equations~(\ref{eq:chi_ode}) and (\ref{eq:Bperp_ode}), are averaged in each bin.

The pixelated proton counts are converted to fluence contrast 
$\Lambda$ using Equation~(\ref{eq:Lambda_def}).  The field reconstruction is 
carried out by solving Equation~(\ref{eq:Recon_eq_diffusion}) for $\phi$ 
numerically on the 128$\times$128 image grid, by the slow-but-effective method 
of unaccelerated Gauss-Seidel iteration (see \S19.5 of \cite{press1992}), which 
gives serviceable results for present purposes. We assume Dirichlet boundary 
conditions, that is, $\phi(\by)=0$ at the image boundary. This scheme, which in 
effect solves a linear problem of the form $\bA\bx=\bb$, iteratively reduces 
the residual $|\bA\bx-\bb|$, where $||$ refers to the 2-norm. We monitor the 
normalized L2-norm of the residual, which is $|\bA\bx-\bb|/|\bb|$, for 
convergence.

The initial guess is a solution of the 
Poisson equation obtained from Equation~(\ref{eq:Recon_eq_3}) by setting the 
deflection $\nabla\phi=0$ in the argument of $\Lambda$ on the RHS. This 
solution can be obtained very rapidly by FFT techniques, but yields a poor 
match to the true field, and we therefore only use it to initialize the 
iterative scheme.

The deflections, and the field projection, are then recovered using 
Equations~ (\ref{eq:w_irrotational}) and (\ref{eq:Recon_B}).

\subsection{An Isolated Blob}\label{subsec:1blob}

We begin with an easy, clean-room case: an ellipsoidal blob, whose image is 
well-separated from the boundaries of the image.  The field strngth 
$B_0=2\times 10^4$~G, and the azimuthal symmetry axis of the blob is aligned 
with the $z$-axis.  The blob has geometric parameters $a=0.03$~cm, $b=0.02$~cm, 
and is centered on the $z$-axis (that is, down the middle of the beam).

In this 
test, we fire 10 million protons at the circular target, and after collimation 
to the square image we have a yield of 6.1 million protons.

In the top row of Figure~\ref{figure:base_case_current}, we show the comparison 
of the nonlinear fluence contrast $\Lambda$ computed from the pixelated data 
(left panel) to the current projection function $\chi$, integrated along proton 
trajectories (middle panel).  The agreement is visually very good.  To perform 
a more quantitative comparison, in the right hand panel we show the normalized 
difference $(\Lambda-\chi)/\sigma_\Lambda$, where $\sigma_\Lambda$ is the 
per-pixel Gaussian error.  This is a reasonable procedure, since the mean 
counts per pixel in the simulation is 373, so that the pixel count distribution 
is well-described by Gaussian statistics. The error $\sigma_\Lambda$ for a 
pixel enclosing $n$ protons is computed starting from the Poisson error for the 
counts, $\sigma_P=\sqrt{n}$, and using the standard propagation of errors 
formula the functional relationship between $\Lambda$ and $n$.  In terms of the 
average counts per bin, $n_0$, we have
\beq
\sigma_\Lambda=\frac{\sqrt{n}}{n_0}.\label{eq:sigma_L}
\eeq
We can see from the top-right panel of Figure~\ref{figure:base_case_current}
that the current projection function predicts the fluence contrast perfectly.
The fluctuations shown in the figure may further be summed in quadrature, to
supply a $\chi^2$ value of 16585.1 for 16384 degrees of freedom (DOF).  This
corresponds to a $P$-value of 13.4\%, indicating a perfectly acceptable 
(parameter-free) ``fit'' of the model to the data.

In the lower panels of Figure~\ref{figure:base_case_current}, we show analogous 
results for the same simulations, using raw counts and the predicted counts per 
pixel.  Once again, the comparison is visually and quantitatively excellent.

The reconstruction of the field configuration is displayed in the upper panels 
of Figure~\ref{figure:base_case_reconstruction}. We ran the iterative solver 
for 4000 iterations, at which point the residual error was 0.51\%. The top-left 
panel shows the actual field projection $l_i\bar{\bB}_\perp$ from the 
integration  of Equation~(\ref{eq:Bperp_ode}), while the top-right panel shows 
the reconstructed field.  The colormap displays field strength on a logarithmic 
scale. We see that the field reconstruction in the center of the image, near 
the peak field-strength location, is very good, while a good deal of 
unstructured noise appears in portions of the image where the field strength is 
more than three orders of magnitude less than the peak value.  At the location
of peak field strength, the difference between the actual and reconstructed field
strengths is 2.0\%. A quantitative 
comparison of the two images may be performed using the $L_2$ norm,
\beq
L_2\equiv\left[
\frac{\sum_{p\in\mbox{pixels}}|\bB^{(\mbox{\scriptsize 
Reconstructed})}_p-\bB^{({\mbox{\scriptsize True})}}_p|^2}
{\sum_{p\in\mbox{pixels}}|\bB^{(\mbox{\scriptsize True})}_p|^2}.
\right]^{1/2}
\eeq

For this reconstruction, we find $L_2=0.13$.  The discrepancy is presumably a
composite of Poisson noise (itself around 5\% per pixel), discretization error,
and algorithmic error (the algorithm terminated with a 0.53\% residual).

The lower panels in Figure~\ref{figure:base_case_reconstruction} show the 
concomitant proton deflections, as reconstructed by the algorithm.  These 
figures illustrate the expected focusing effect of the azimuthal field.  Again,
the vector displacements maintain good accuracy in the range from peak 
displacement length to displacements that are about three orders of magnitude 
down from peak, after which unstructured noise appears.

\subsection{Two Superposed Blobs}\label{subsec:twoblobs}

We also simulate a more complex, and less symmetric situation, by superposing
two blobs of unequal strength -- one at $y=+0.025$~cm, with $B_0=2\times 
10^4$~G, and one at $y=-0.025$~cm, with $B_0=10^4$~G.  The geometric parameters
of the blobs are both identical to those in the previous example -- 
$a=0.03$~cm, $b=0.02$~cm.  We again fire 10 million protons at the system, 
recovering 6.1 million protons after collimation to the square image.

Comparisons of fluence contrast to current projection, and of counts per pixel 
to predicted counts per pixel, are displayed in 
Figure~\ref{figure:twoblob_current}.  The comparison is again both visually and 
quantitatively satisfactory.  The difference of the fluence contrast and 
current projection maps yields a $\chi^2=16448.4$ for 16384 DOF, a $P$-value of 
36\%.  The model is quite clearly in good agreement with the data.

The reconstruction is illustrated in 
Figure~\ref{figure:twoblob_reconstruction}.  The iterative solution residual 
after 4000 iterations is 0.60\%. Again, the visual impression of the 
reconstructed field is quite good, with good structural fidelity within three 
orders of magnitude peak, for both magnetic field and for deflections. At the 
peak field strength location, the difference between the reconstructed and 
actual field strengths is 2.1\%. The L2 norm of the difference between the reconstructed 
and actual fields is $L_2=0.13$.

\section{Exploration and Characterization of Reconstruction Errors}\label{sec:errors}

\begin{figure*}[t]
\begin{center}
\includegraphics[width=6.0in, viewport=90 70 1140 940, clip=]{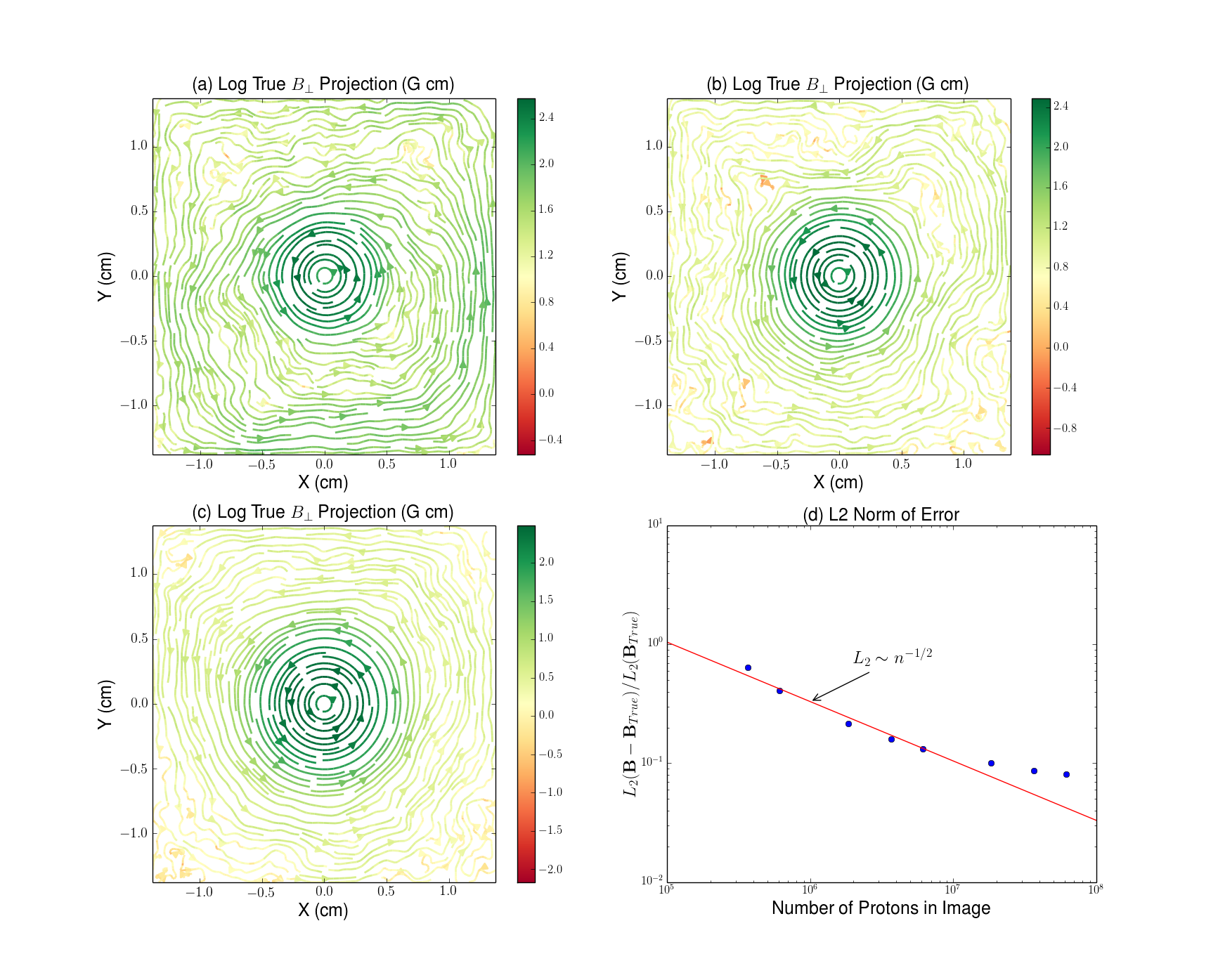}
\end{center}
\caption{\label{Figure:Poisson}Study of the effect of Poisson noise. 
The field reconstruction figures display the field as reconstructed
for different numbers of protons in the field of view -- $3.7\times 
10^5$ (a), $3.7\times 10^6$ (b), and 
$6.1\times 10^7$ (c).  Panel (d) displays
the error of the reconstruction, measured using the relative $L_2$ 
norm of the difference between the reconstructed field and the true 
field.  The line shows the expected $n^{-1/2}$ scaling characteristic 
of Poisson noise errors.  For these simulations, the Poisson noise 
error dominates until $n\approx 10^7$, at which point other 
types of error take over the error budget.}
\end{figure*}

In this section, we explore the character of some errors that affect the 
analysis of radiographs.  We are concerned here not with the general subject of 
systematic errors in proton radiography (a subject that received a very 
thorough treatment in K2012), but rather with errors that limit the accuracy of 
the magnetic field reconstruction.

\begin{figure*}[t]
\begin{center}
\includegraphics[width=6.8in, viewport=145 0 1400 425, clip=]{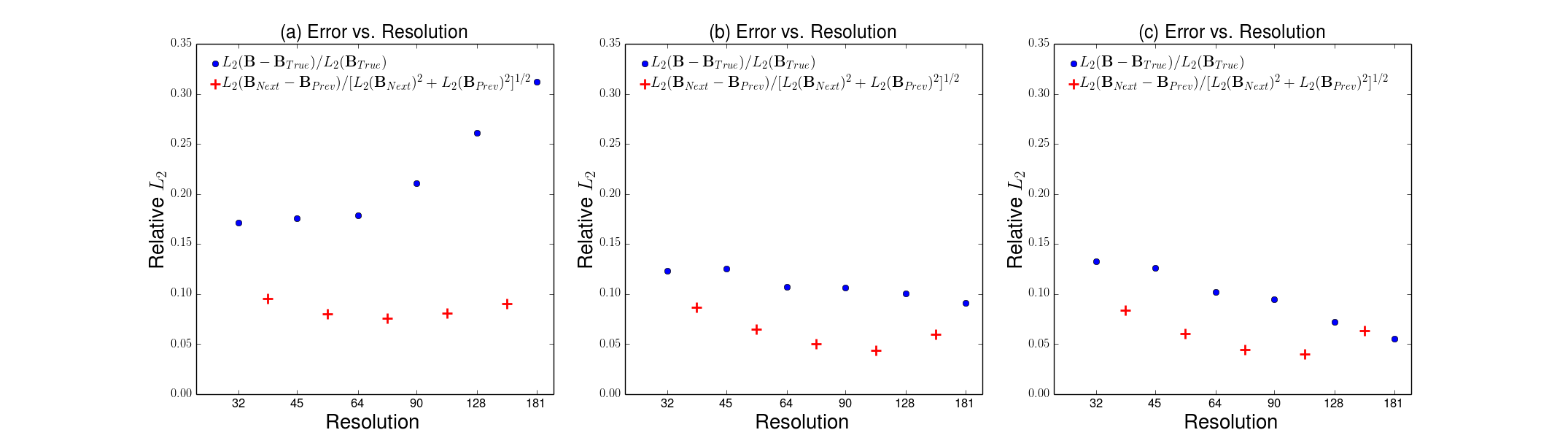}
\caption{\label{figure:resolution} Resolution study. The three figures show relative
$L_2$ norm measures for three different proton fluences.  (a) 
$6.1\times 10^5$ protons in the square FOV; (b) $2.4\times 10^6$ 
protons in FOV; (c) $6.1\times 10^6$ protons in FOV.  The blue dots 
show the $L_2$ measure of departure of the reconstructed field from the true 
field, at each of a number of grid resolutions (number of mesh points per 
dimension). The red crosses show the $L_2$ difference between the 
reconstructions corresponding to the two adjoining resolutions.  The turnover 
point in the crosses can serve as an indicator of ``optimal'' (or, at least, 
not horrible) resolution in experimental work, where the ``true'' field is not 
available for comparison.}
\end{center}
\end{figure*}

We can identify several individual (although not necessarily 
independent) sources of error that are relevant to reconstruction:

\begin{description}

\item{Poisson Noise:} In the diffusion equation that we use for 
reconstruction, Equation~(\ref{eq:Recon_eq_diffusion}), both the 
source term on the right-hand side and the diffusion coefficient are 
estimated noisily from binned proton counts.  This Poisson noise is 
a source of error whose principal effect is uncertainty that scales 
as $n^{-1/2}$, where $n$ is the average number of protons per pixel. 
However, we shall see that Poisson noise can also be expected to 
produce other indirect effects;

\item{Discretization Noise:} The numerical solution is carried out on
a uniform grid in the image plane.  Obviously, the level of refinement 
of the grid has an impact on the accuracy of the solution.  In 
principle, the mesh should be fine enough to resolve the smallest 
magnetic structures in the image. However, 
one may not simply refine the mesh indefinitely, as the reduced error 
due to discretization will eventually be balanced by Poisson noise as 
the average number of protons per pixel drops, and the form of the 
fluence contrast function $\Lambda$ becomes noisily determined.  Thus, 
there is a tradeoff between discretization noise and Poisson noise 
that should be optimized somehow;

\item{Edge Effects:} The solution of an elliptical PDE such as 
Equation~(\ref{eq:Recon_eq_diffusion}) requires assumptions about boundary 
condiditions.  A natural choice is Dirichlet BC, with the function 
$\phi(\by)\rightarrow 0$ at the edges of the radiograph. If the active region 
is well-separated from the edges, then this choice may be expected to give good 
results.  If the active region is \textit{not} well-separated from the edges, 
this choice is not ideal, and may at best be regarded as an approximation of 
convenience.  In this case, it seems clear that the ability to recover detailed 
field structure is compromised. It is interesting to inquire whether one may 
still infer the correct order of magnitude of $|\bB_\perp|$ in these 
circumstances. It should be also be noted that even for an active region that 
is isolated from the boundaries, Poisson noise can furnish spurious activity 
near the boundary.  Thus, edge effects and Poisson noise effects are also 
intertwined;

\item{Obstruction Effects:} It is not uncommon in laser experiments 
for some part of the physical structure of the target to be 
illuminated by the beam of protons, and thus to cast a shadow on the 
image.  Such a shadow, treated naively, would appear to the present 
analysis as an enormous, proton-clearing field of large strength and 
strong gradients! Evidently it is necessary to characterize the 
importance of this kind of error. One very crude fix is to replace 
the proton fluence in the shadowed regions by the mean, 
zero-deflection proton fluence.  If the shadow encroaches on an image 
region of magnetic activity, this procedure will clearly spoil the 
reconstruction of the detailed field structure.  We may still hope to 
at least recover the typical field strengths, at least approximately.  

\item{The Diffusion Approximation:} The linear diffusion PDE in 
Equation~(\ref{eq:Recon_eq_diffusion}) is an approximation to the 
non-linear reconstruction PDE in Equation~(\ref{eq:Recon_eq_3}).  The 
approximation is valid if we may legitimately expand the RHS of 
Equation~(\ref{eq:Recon_eq_3}) to linear order in $\nabla\phi$.  This
means that the displacement field $\nabla\phi$ must be ``small'' in a
different sense than the previous condition $\alpha\ll 1$:  it must 
also be true that the second-order term in the Taylor expansion of 
the RHS of Equation~(\ref{eq:Recon_eq_3}) should be small compared to
the linear term, that is (with apologies for the abuse of notation) 
that $|\nabla\phi|\ll|\nabla\Lambda|/|\nabla\nabla\Lambda|$, everywhere
on the image.

\end{description}

Below, we explore numerically some of the characteristics of Poisson 
noise, discretization noise, edge effects, and obstruction effects. 
We will overlook the 
effect of the diffusion approximation, since it would require a 
substantial investigation, if for no other reason that one would 
first have to devise some method of solving the non-linear PDE in 
Equation~(\ref{eq:Recon_eq_3}). In our opinion, the effort involved 
justifies a separate paper.

\subsection{Poisson Noise Effects}

We explore the influence of Poisson noise using the first field 
simulated in the previous section, which was displayed in Figure~\ref
{figure:base_case_reconstruction}.  We shoot $10^8$ protons at this 
field, and produce datasets of different fluences by truncating this 
simulation to $6\times 10^5$, $10^6$, $3\times 10^6$, $6\times 10^6$,
$10^7$, $3\times 10^7$, $6\times 10^7$ and $10^8$ protons.  In each 
case, after collimation to the square field-of-view, about 60\% of the 
protons remain.  We perform the reconstruction procedure on each 
dataset, using the same $128\times 128$ grid that we used in the 
previous section.

Three sample reconstructions are displayed in 
Figure~\ref{Figure:Poisson}. They correspond to the datasets with 
$6\times 10^5$ protons ($3.7\times 10^5$ in the FOV), $6\times 10^7$ 
protons ($3.7\times 10^7$ in the FOV), and $10^8$ protons ($6.1\times 
10^7$ in the FOV).  The progressive degradation of the reconstruction 
is evident, and proceeds from the regions of weaker field (where the 
current integral is more poorly determined) inwards to the regions of 
stronger field.

The lower-right panel shows the $L_2$-norm of the difference between 
the reconstructed field and the actual field, normalized to the 
$L_2$-norm of the actual field, as a function of number of protons in 
the FOV.  The error shows the expected Poisson scaling of $n^{-1/2}$ 
for $n\lesssim 10^7$.  Evidently, for these simulations, other 
types of error begin to dominate the error budget above this proton 
fluence.

\subsection{Discretization Effects}

As discussed above, the effect of grid spacing on reconstruction accuracy is
inextricably entwined with the effect of Poisson noise.  For a fixed 
radiograph, the resolution can only be increased up to the point where the 
per-pixel proton counts begin to be so low that Poisson noise begins to degrade 
the reconstruction.

\begin{figure*}[t]
\begin{center}
\includegraphics[width=6.5in, viewport= 90 20 1100 425]{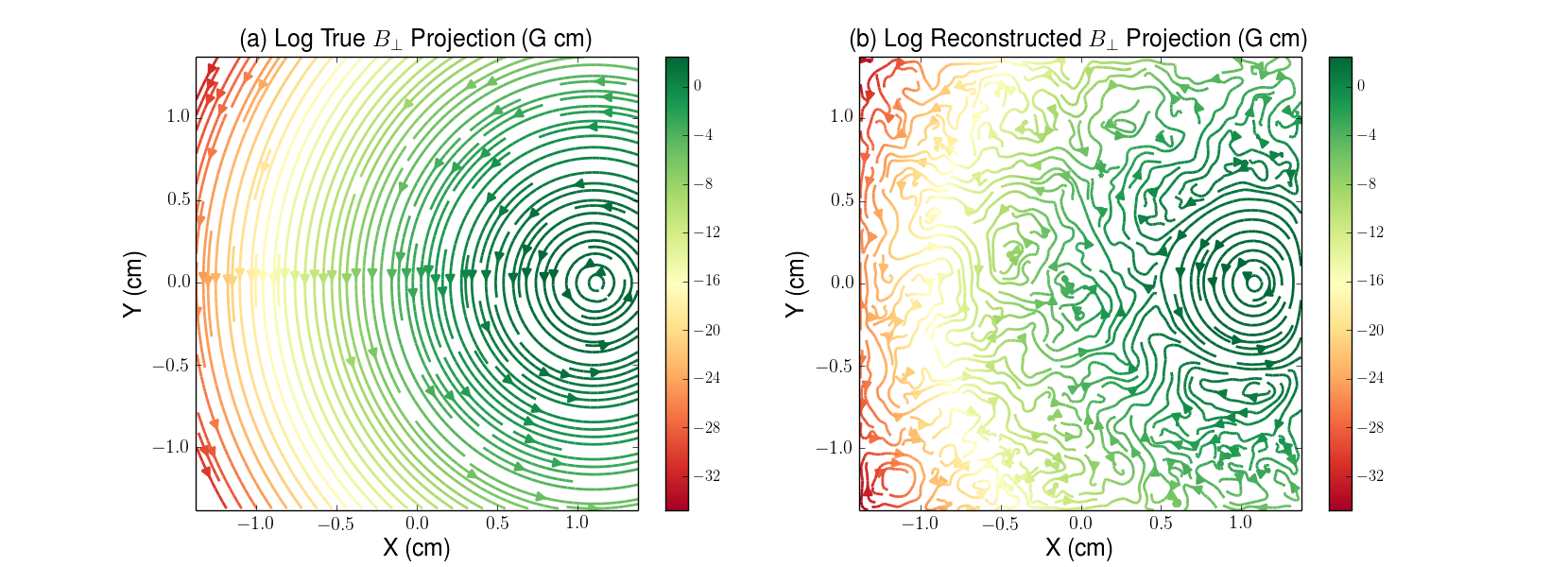}
\caption{\label{figure:edge_reconstruction} Reconstructed field configuration
for a blob situated near the edge of the imaged region.}
\end{center}
\end{figure*}

\begin{figure*}[t]
\begin{center}
\includegraphics[width=6.8in, viewport=175 0 1660 425, clip=]{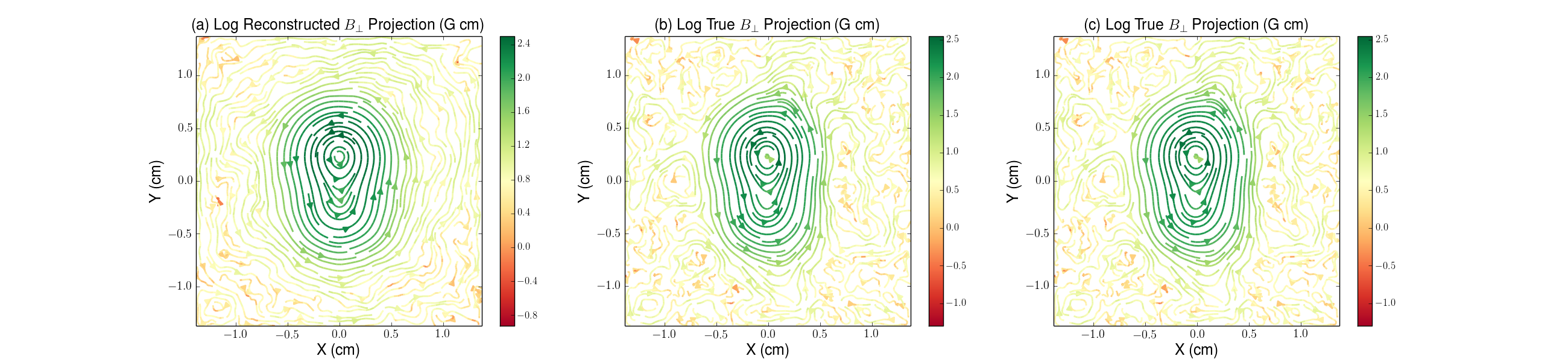}
\caption{\label{figure:obstruction} Obstruction study. The three figures
show reconstructed magnetic field for the two-blob test case, similar
to Figure~\ref{figure:twoblob_reconstruction}, but in each case with
a strip of proton data removed and replaced by the overall average 
proton flux.  (a) No obstruction (same as in Figure~\ref{figure:twoblob_reconstruction}.  (b)
Data obstructed in the vertical strip $0.1<x<0.2$.  (c) Data 
obstructed in the vertical strip $0.1<x<0.5$.
}
\end{center}
\end{figure*}

To explore the connection between resolution and Poisson noise, we conducted
three series of reconstructions.  Each series comprised a set of six grids -- 
$32\times32$, $45\times 45$, $64\times 64$, $90\times 90$, $128\times 128$, and
$181\times 181$ grid points, respectively (each contains twice as many points 
as the previous one).  We performed this series of reconstructions on each of 
three simulations of a blob.  The three simulations differed only in the number 
of protons -- they contained $6.1\times 10^5$, $2.4\times 10^6$, and $6.1\times 
10^6$ protons in the square FOV, respectively.  The selected blob had 
parameters that differed from the one in \S\ref{subsec:1blob}, in the single 
respect that $a=0.05$~cm instead of $a=0.03$~cm.  We expanded the blob 
laterally so as to give the reconstructions a better chance to capture the 
field at the low resolution end.

The results are displayed in Figure~\ref{figure:resolution}.The blue circles 
show the relative $L_2$ norm of the difference between the reconstructeed field 
and the ``true'' field. The red crosses show the relative $L_2$ norm of the 
difference between two ``adjacent'' resolutions (the crosses are therefore 
located between their corresponding pairs of circles). The value of the latter 
measure is that it is available in experimental situations, when the true field 
is not known.

The figure shows the very strong effect of proton fluence on reconstruction 
accuracy.  Depending on the proton fluence, the reconstruction accuracy may be 
compromised even at low resolutions, or alternatively it may continue to 
improve up to the limit of our patience with the reconstruction algorithm 
(which has a time-to-solution that grows at least as fast as the number of grid 
points).  The difference between adjacent resolutions appears to furnish a 
reasonable ``rough-and-ready'' indicator of optimality (or, at least, of 
non-pessimality), in the sense that in the low-fluence case (left panel) the 
crosses begin to rise at about the time when the departure from the true field 
begins to get serious.

\subsection{Edge Effects}

To illustrate the influence of boundary effects on field reconstruction we 
perform a simulation using a blob with the same parameters as the ones used in 
\S\ref{subsec:1blob}, but displaced so that a substantial part 
of the previously-reconstructed field now overlaps the right edge of the image. 
To do this, we displace the interaction region to the right by 1.1$l_i$.  We 
shoot $10^7$ protons at the target as before, and again recover about $6\times 
10^6$ protons in the collimated square image.

The resulting reconstruction is displayed in Figure~\ref
{figure:edge_reconstruction}.  The comparison with the single 
isolated blob of \S~\ref{subsec:1blob} is instructive.  In this 
case, the detailed reconstruction accuracy is degraded -- the $L_2$ 
norm of the difference between actual and reconstructed fields has 
increased to 29\% (from 13\%), and the difference between actual and 
reconstructed field strengths at the location of peak field strength 
has increased to 13.5\% (from 2.0\%).  On the other hand, the 
structure of the field is still recognizable:  the reconstructed 
contours are still clearly concentric circles to a radius of 0.5~cm 
from the center point of the field, which was also the case in 
Figure~\ref{figure:base_case_reconstruction}.  It appears, then, 
that when the image of the interaction region encroaches on the edge 
of the radiograph, one may still hope to infer field strength peak 
magnitudes and field structure to some useful degree, despite the 
inexactly obeyed boundary conditions.

\subsection{Obstruction Effects}

We study the effect of obstruction on field reconstruction using a 
variation of the two superposed blob simulation described in 
\S\ref{subsec:twoblobs}, where we summarily remove the protons in a 
vertical strip near the blob, and replace the removed fluences by the 
mean, zero-deflection fluence for the image.  We do this for 
obstruction strips of two different widths.  

The results are shown in Figure~\ref{figure:obstruction}.  The left 
panel of this figure shows the result of reconstruction using 
unobstructed data, and is therefore the same as the reconstruction 
in Figure~\ref{figure:twoblob_reconstruction}.  The middle panel 
shows the reconstruction that is obtained after obstructing the data 
in the vertical strip $0.1<x<0.2$, while the right panel shows the 
reconstruction obtained after obstructing data in the strip 
$0.1<x<0.5$.

What emerges from this study is that the field reconstruction can be 
remarkably robust to obstruction effects.  In both obstructed cases, 
the typical field intensities are largely unaffected by the 
obstruction, and the morphology is not wildly distorted.  This is, 
we believe, due to the nonlocal nature of the reconstruction, which 
can partly ``heal'' the data corruption, interpolating to preserve 
the assumption of continuity and differentiability of the field that 
is built in to the reconstruction procedure.  Naturally, one would 
expect the fidelity of the reconstruction  to be more seriously 
compromised as the obstruction becomes more serious.  It is also 
likely that the unrealistic simplicity of the two-blob test case is 
helpful in aiding reconstruction.  Nonetheless, this test case offers 
some reason to place some limited reliance on field strengths and 
morphologies inferred through the reconstruction procedure, even in 
the case of obstructed data.

\begin{figure*}[t]
\begin{center}
\includegraphics[width=6.8in, viewport=95 0 1090 425, clip=]{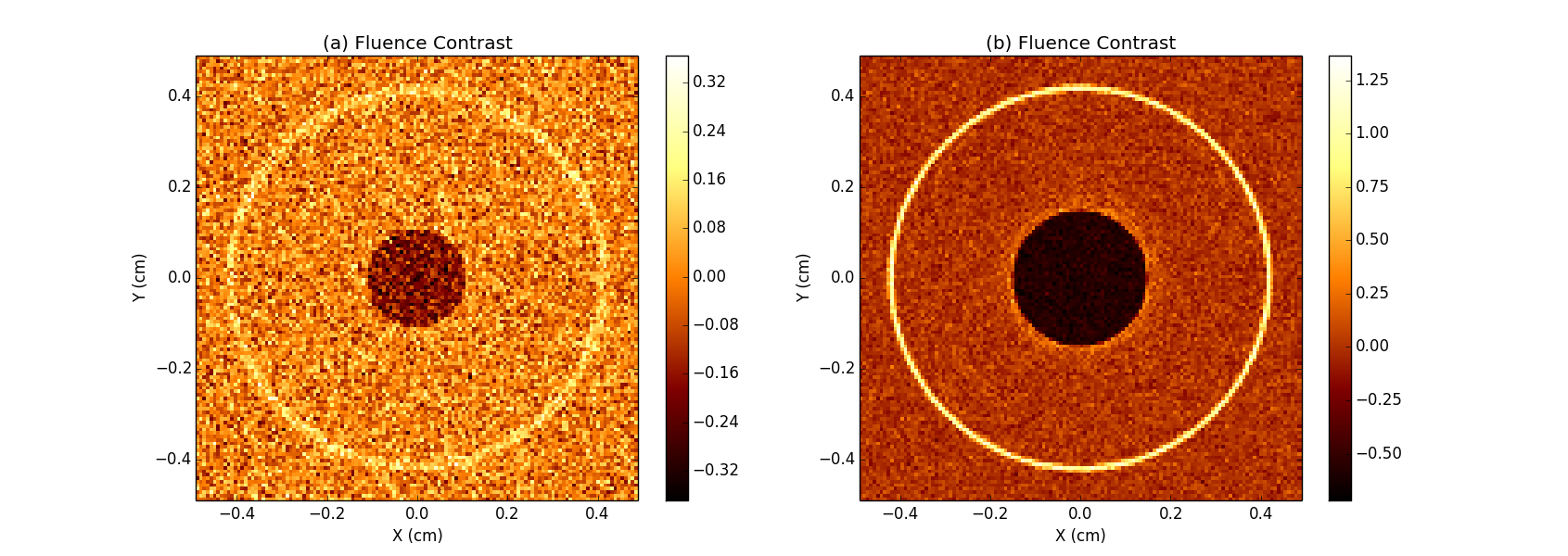}
\caption{\label{figure:beercan}``Wire in a beer can'' simulation. A uniform current 
runs along a central wire, with the return current flowing back down the walls 
of the can. (a) Linear contrast regime.  (b) Nonlinear contrast
regime.}
\end{center}
\end{figure*}

\subsection{The Trouble With Caustics}\label{subsec:Caustics}

The theory developed in this article assumes linear contrast departures, 
which, as we have seen, means small field gradients, and, particularly, small 
currents.  In this regime, there is no strong focusing, and the caustic 
structures examined and described in K2012 cannot develop in 
radiographic images.  Nonetheless, it is noteworthy that high-definition 
structures \textit{can} develop in images, even in the absence of caustic 
structure. In fact it is apparent that \textit{the appearance of sharp features 
in a proton radiograph cannot be taken as evidence of caustic formation}.  
Furthermore, as we have seen the essential tool for understanding radiographic 
structure is not caustic structure, but rather the MHD current.

As an aid to the discussion, consider the ``wire-in-a-beer-can'' field 
configuration, wherein a uniform current flows along a central wire of finite 
radius, and the return current flows back along the finite-thickness walls of 
the can. The wire diameter is $r_{0}$, the can diameter is $r_{1}$, the 
thickness of the can walls is $\delta$, and the height of the can is $z_{0}$. 
The net current in the central wire is $I$. We assume the current density is 
uniform in the wire and in the can's vertical side walls.

The magnetic field $\bB$ is purely azimuthal, $\bB=B_{\phi}\mathbf{e}_{\phi}$, 
and we assume azimuthal symmetry, so that $\nabla\cdot\bB=r^{-1}\partial 
B_{\phi}/\partial\phi=0$.

We choose a factorized form for $B_{\phi}$:
\bea
B_{\phi}=f(r)g(z),\label{eq:factorization}
\eea
where $r$ is the cylindrical radius coordinate, and $z$ is cylindrical
height coordinate.

The $z$-dependence of the field simply limits the field to the can
vertically:
\bea
g(z)=\begin{cases}
1 & ,\,|z|<z_{0}/2;\\
1-\frac{|z|-z_{0}/2}{\delta} & ,\,z_{0}/2\le|z|<z_{0}/2+\delta;\\
0 & ,\,|z|\ge z_{0}/2+\delta.
\end{cases}\label{eq:Vertical}
\eea

The function $f(r)$ is set by the can/wire geometry described above,
and by Ampere's law
\bea
B_{\phi}=\frac{2i(r)}{cr},\label{eq:Ampere_Law}
\eea
where $i(r)$ is the total net current along the $z$-direction inside
radius $r$. By considering the region $|z|<z_{0}/2$ we therefore
have
\begin{equation}
f(r)=\begin{cases}
\frac{2Ir}{cr_{0}^{\,2}} & ,\,r<r_{0};\\
\frac{2I}{cr} & ,\,r_{0}\le r<r_{1};\\
\frac{2I}{cr_{i}}\left(1-\frac{r-r_{i}}{\delta}\right) & ,\,r_{1}\le r<r_{1}+\delta;\\
0 & ,\,r\ge r_{i}+\delta.
\end{cases}\label{eq:Radial}
\end{equation}
This field configuration yields a uniform current density in the wire
and in the vertical walls of the can.
The current $\mathbf{J}$ bends to flow around the end-caps from
the wire to the side walls and back. The azimuthal magnetic field strength rises 
linearly from the center to the edge of the wire, then declines as $r^{-1}$ in 
the interior of the can, and finally drops linearly to zero in the outer wall 
of the can.
 
Figure~\ref{figure:beercan} gives two simulated (with 5 million protons) 
radiographic images of ``wire-in-a-beer-can'' simulations. The current 
projection is negative in the central wire, positive in the walls. In the left 
panel, the parameters are chosen so that the fluctuations are linear -- the 
contrast is everywhere less than about 0.3. In the right panel, the field 
intensity has been increased to push the contrasts well into the nonlinear 
regime, so that the central region is nearly cleared of protons, and the walls 
have more than double the mean count rate.

There is no question of any caustic structure in the first image, since the 
contrasts are too small for the necessary nonlinear structure to be present. In 
the second image, it is likely that a caustic has developed somewhere in the 
outer ring, given the high constrast there.  Nonetheless we see very similar 
morphology in the two cases.  This illustrates the point that caustic 
formation is in no sense necessary for the formation of sharp radiographic 
structure.  We conclude that it is not in general correct to identify 
high-definition features in radiographs with caustic structure, in an effort 
to infer magnetic structure using the formulae in K2012 to relate such 
structure to image caustics.  At a minimum, one should verify the necessary 
condition for caustic formation -- large contrast fluctuations.

The importance of using MHD current to interpret radiographs can be seen by 
noting that intuitive interpretations of the field configuration that do not 
rely on the behavior of the current can be seriously misleading.  In 
particular, it would be incorrect to associate the peak field intensity with 
the peak count rate, since according to Equation~(\ref{eq:Radial}) the field 
strength peaks at the surface of the wire, rather than at the side walls of the 
can.  A somewhat less naive interpretation would be to associate field 
gradients only with the interiors of the images of the wire (the central hole) 
and the wall (the outer annulus).  This would also be incorrect, however, since 
according to Equation~(\ref{eq:Radial}), $|df/dr|$ is also nonzero in the 
interior of the can ($r_0\le r<r_1$), where there is no image contrast at all! 
Furthermore, on the basis of either image, one might naively ascribe the same 
(zero) field to the exterior of the can ($r>r_1+\delta$) as to the interior 
($r_0\le r<r_1$), since the images of both regions are uniform-no-contrast.  In 
fact, the interior of the can contains a non-zero irrotational field, while the 
exterior field is zero.

All these features are perfectly intelligible when the current is considered. 
The current along the proton beam direction is negative in the central wire, 
positive in the walls, and zero elsewhere.  That statement gives a completely 
satisfactory and concise description of both the linear and the nonlinear 
image.

A further defect of the emphasis on caustic structure is that it would lead the 
analyst to emphasize the bright annulus of the wall, where the strong focusing 
occurs, over the hole in the center, where strong \textit{de-focusing} is the 
story. But the two are complementary, as can be seen by considering how the 
image would change were the current simply reversed everywhere -- the central 
hole would become a bright spot (and presumably the locus of a caustic in the 
nonlinear image), the wall a trough.  The complementarity between the two features, 
which is obvious when their interpretation is in terms of current, is lost when 
one adopts instead the caustic as one's interpretive key. 

This leads to a final point.  It is \textit{current}, not caustic structure, 
that makes these images intelligible.  Even in the non-linear regime, despite 
the invalidation of the linear theory, we see that current continues to be a 
valuable guide to image structure.  This consideration is important for image 
analysis, but vital for experimental design:  the decision on where 
to position and how to orient a radiographic setup should be made, to the 
maximum extent possible, on the basis of the morphology of the MHD currents 
that one expects to be produced in the experiment.  Setting up the proton beam 
largely perpendicularly to expected current flows would result in avoidably 
weak signal; contrariwise, the signal may be optimized by aligning beam 
and currents to the greatest extent possible.  This highlights the importance 
of high-fidelity simulations of laser plasma experiments, such as the ones 
described in \textcite{tzeferacos2015}.  Such simulations easily allow one to 
discern the main current flows in the plasma, and hence optimize
one's radiographic setup with respect to the experimental setup.

\section{Proton Radiography of Isotropic-Homogeneous Turbulent Fields}
\label{sec:turbulence}

In experimental applications that result in turbulent flows, the direct 
reconstruction of the transverse magnetic field is not necessarily desirable. 
Of greater interest is the spectrum of the turbulent magnetic field.  While the 
spectrum is in principle recoverable from the reconstructed field, such a 
procedure would unnecessarily entangle the inferred spectrum with some of the 
reconstruction errors discussed above (i.e. edge and obstruction effects) As we 
will now demonstrate, under the assumptions that the field has an approximately 
uniform and isotropic spectrum, the spectrum itself is directly and robustly 
recoverable from the radiographic image, without performing the field 
reconstruction.

Our starting point is Equation~(\ref{eq:dw_2}), which we rewrite slightly as
\bea
\Lambda(\bx_\perp)=
\frac{er_i(r_s-r_i)}{mcvr_s}\hat{\bz}\cdot\int_{r_i-l_i/2}^{r_i+l_i/2}dz\,
\nabla\btimes\bB(\br(z,\bx_\perp)),
\label{eq:Psi3}
\eea
where we accept to incur an error of order ${\cal O}(\omega)$ by replacing 
$\bn^{(0)}$ with $\hat{\bz}$ in the inner product, and one of order 
${\cal O}(\lambda)$ by separating the  dependence on $z$ from that on $\bx_\perp$ 
in the argument of the integrand, setting
\beq
\br(z,\bx_\perp)\equiv z\hat{\bz}+\frac{r_i}{r_s}\bx_\perp.
\label{eq:depsep}
\eeq
In Equation~(\ref{eq:Psi3}), we have also changed the limits of integration to 
include only the region in which the magnetic field is appreciable.  The reason 
for this change will be made apparent below.

Note also that in Equation~(\ref{eq:Psi3}) we are neglecting the small shift in the 
argument of $\Lambda$ that was so important to the reconstruction algorithm 
(compare Equation~\ref{eq:Recon_eq_2}).  This is permissible at the cost of 
stipulating that the spectrum obtained thereby cannot be trusted at scales 
close to, or smaller than, the angular deflection scale $\alpha l_i$. 
At longer scales, the spectrum should be unaffected by this approximation.

\subsection{The Two-Point Correlation Function and the Spectrum}

We define the two-point correlation function for the radiographic image as 
follows:
\beq
\eta(\left|\mathbf{x}_{\perp}^{\prime}-\mathbf{x}_{\perp}\right|) \equiv 
\left\langle
\Lambda(\bx_\perp)\Lambda(\bx_\perp^\prime)
\right\rangle,
\label{eq:tpfdef}
\eeq
where  the expectation value $\left\langle \right\rangle $ is an ensemble 
average. The fact that 
$\eta\left(\left|\mathbf{x}_{\perp}^{\prime}-\mathbf{x}_{\perp}\right|\right)$ 
depends only on the distance between $\mathbf{x}_{\perp}^{\prime}$ and 
$\mathbf{x}_{\perp}$ is a consequence of the assumed isotropy and homogeneity 
of the stochastic field $\mathbf{B}$, as will be shortly evident.

Inserting Equation~(\ref{eq:Psi3}) in Equation~(\ref{eq:tpfdef}) we get
\bea
&&\left\langle
\Lambda(\bx_\perp)\Lambda(\bx_\perp^\prime)
\right\rangle =\nonumber\\
&&\hspace{1cm}\frac{e^2r_i^2(r_s-r_i)^2}{m^2c^2v^2r_s^2}\epsilon_{lmn}
\epsilon_{l^{\prime}m^{\prime}n^{\prime}}\hat{z}_{l}\hat{z}_{l^{\prime}}\nonumber \\
&&\hspace{1cm}\times
\int_{r_{i}-l_{i}/2}^{r_{i}+l_i+1/2}dz
\int_{r_{i}-l_{i}/2}^{r_{1}+l_{i}/2}dz^{\prime}\nonumber\\
&&\hspace{1cm}
\frac{\partial}{\partial r_{m}}\frac{\partial}{\partial r_{m^{\prime}}^{\prime}}
\left\langle 
B_{n}\left(\br(z,\bx_{\perp})\right)
B_{n^{\prime}}\left(\br(z^{\prime},\bx_{\perp}^{\prime}\right)
\right\rangle.
\label{eq:Correlation_1}
\eea

The correlation expression in the integrand defines an isotropic tensor
\beq
\left\langle B_{n}\left(\bx\right)B_{n^{\prime}}\left(\bx^{\prime}\right)
\right\rangle 
\equiv Q_{nn^{\prime}}\left(\bx-\bx^{\prime}\right)\label{eq:QT}
\eeq
that is analogous to the classic incompressible velocity correlation
tensor, since $\nabla\cdot\mathbf{B}=0$ is analogous to the incompressibility
condition for velocity. Following the development in Sections 6.2
and 8.1 of \textcite{davidson2004turbulence}, we write
\beq
Q_{nn^{\prime}}(\ba)=\int d^{3}\bk\,e^{i\bk\cdot\ba}\Phi_{nn^{\prime}}(\bk),\label{eq:QTensor}
\eeq
where the isotropic tensor $\Phi_{nn^{\prime}}(\bk)$ has the
form
\beq
\Phi_{nn^{\prime}}(\bk)=\frac{E_{B}(k)}{k^{2}}\left(
\delta_{nn^{\prime}}-k_{n}k_{n^{\prime}}/k^{2}
\right).\label{eq:PhiTensor}
\eeq
The quantity $E_{B}(k)$ is the energy spectrum of the magnetic field,
and satisfies the relation
\beq
\int_{0}^{\infty}dk\,E_{B}(k)=\frac{1}{8\pi}\left\langle \bB^{2}\right\rangle,
\label{eq:Espectrum}
\eeq
which is obtainable by setting $\ba=0$ and contracting the indices in 
Equation~(\ref{eq:QTensor}).

Inserting Equations~(\ref{eq:QT}-\ref{eq:PhiTensor}) into Equation~(\ref{eq:Correlation_1}),
we obtain
\bea
&&\left\langle
\Lambda(\bx_\perp)\Lambda(\bx_\perp^\prime)
\right\rangle =\nonumber\\
&&\hspace{1cm}\frac{e^2r_i^2(r_s-r_i)^2}{m^2c^2v^2r_s^2}\epsilon_{lmn}
\epsilon_{l^{\prime}m^{\prime}n^{\prime}}\hat{z}_{l}\hat{z}_{l^{\prime}}\nonumber \\
&&\hspace{1cm}\times
\int_{r_{i}-l_{i}/2}^{r_{i}+l_i+1/2}dz
\int_{r_{i}-l_{i}/2}^{z_{1}+l_{i}/2}dz^{\prime}\nonumber\\
&&\hspace{1.2cm}\int d^{3}\bk\,
e^{i\bk\cdot\left(\br(z,\bx_{\perp})-\br(z^{\prime},\bx_{\perp}^{\prime})\right)}\nonumber\\
&&\hspace{1.2cm}\times k_{m}k_{m^{\prime}}
\frac{E_{B}(k)}{k^{2}}
\left(\delta_{nn^{\prime}}-k_{n}k_{n^{\prime}}/k^{2}\right).
\label{eq:Correlation_2}
\eea

At this point it is possible to clarify why it was important to 
adopt limits of $z$-integration that are confined to the interaction 
region. The magnetic field is confined to the interaction region, 
and were we to Fourier-transform $\mathbf{B}$ instead of $Q$, the 
various complex phases of the transform of $\mathbf{B}$ would 
contain the information necessary to respect this confinement. 
However, the spectrum $E_{B}(k)$ that appears in the Fourier 
transform of $Q$ knows nothing of those phases. It knows that there 
is an integral scale cutoff $k_{0}\gtrsim2\pi/l_{i}$, but it 
represents a model of uniform, isotropic turbulence in which \emph
{all space }is pervaded by a turbulent $\mathbf{B}$-field with 
integral-scale cutoff $k_{0}$. The restriction on the limits of 
integration in $z$ in effect prevents that model from incorporating 
proton deflections from regions where physically the field is zero.

The term $k_{n}k_{n^{\prime}}$ in the integrand of 
Equation~(\ref{eq:Correlation_2}) vanishes, since $\epsilon_{lmn}k_{m}k_{n}=0$. 
We may then use the decomposition $\bk=\bk_\perp+k_z\hat{\bz}$ and the 
Levi-Civita tensor's contraction identity

\beq
\epsilon_{lmn}\epsilon_{l^{\prime}m^{\prime}n^{\prime}}\delta_{nn^{\prime}}=
\delta_{ll^{\prime}}\delta_{mm^{\prime}}-\delta_{lm^{\prime}}\delta_{ml^{\prime}}
\label{eq:LC_Contraction}
\eeq
to obtain
\bea
&&\left\langle
\Lambda(\bx_\perp)\Lambda(\bx_\perp^\prime)
\right\rangle =\nonumber\\
&&\frac{e^2r_i^2(r_s-r_i)^2}{m^2c^2v^2r_s^2}
\int_{-\infty}^{\infty}dk_z\,
\frac{4\sin^2(k_{z}l_{i}/2)}{k_{z}^{2}}\nonumber \\
&&\times\int d^{2}\bk_{\perp}\,e^{i(r_{i}/r_{s})
\bk_{\perp}\cdot(\bx_{\perp}-\bx_{\perp}^{\prime})}
\frac{E_{B}\left([k_{\perp}^{2}+k_{z}^{2}]^{1/2}\right)k_{\perp}^{2}}
{k_{\perp}^{2}+k_{z}^{2}}.\nonumber\\
\label{eq:Correlation_3}
\eea

As a function of $k_{z}.$ the integrand term 
$4\sin^2(k_{z}l_{i}/2)/k_{z}^{2}$
is sharply peaked near $k_{z}=0$, and has a width $1/l_{i}$. Compared
to the $k$ scales present in $E_{B}(k)$, it is well-approximated
by a $\delta$-function:
\beq
\frac{4\sin^2(k_{z}l_{i}/2)}{k_{z}^{2}}
\approx2\pi l_{i}\delta(k_{z}).
\label{eq:kz-delta}
\eeq

Using this in Equation~(\ref{eq:Correlation_3}), we obtain
\bea
\left\langle
\Lambda(\bx_\perp)\Lambda(\bx_\perp^\prime)
\right\rangle&=&
\frac{2\pi e^2r_i^2(r_s-r_i)^2l_i}{m^2c^2v^2r_s^2}\times\nonumber\\
&&\int d^{2}\bk_{\perp}\,
e^{i(r_{i}/r_{s})\bk_{\perp}\cdot(\bx_{\perp}-\bx_{\perp}^{\prime})}
E_{B}(k_{\perp})\nonumber\\
&=&
\frac{2\pi e^2(r_s-r_i)^2l_i}{m^2c^2v^2}\times\nonumber\\
&&\int d^{2}\bq_{\perp}\,
e^{i\bq_{\perp}\cdot(\bx_{\perp}-\bx_{\perp}^{\prime})}
E_{B}\left(\frac{r_s}{r_i}q_{\perp}\right),\label{eq:Correlation_4}
\eea
where we've defined the screen-plane wave vector 
$\bq_\perp\equiv\frac{ri}{rs}\bk_\perp$.
Since $E_{B}(k_{\perp})$ depends only on the magnitude of $\mathbf{k}_{\perp}$,
we see that the autocorrelation indeed only depends on the magnitude $\left|\mathbf{x}_{\perp}^{\prime}-\mathbf{x}_{\perp}\right|$,
as previously asserted. 

Note that had we not restricted the $z$ integration range, rather leaving it as 
$0<z<z_{s}$, we would have obtained a similar result, but scaled up by 
$r_{s}/l_{i}$ (because instead of the RHS of Equation~(\ref{eq:kz-delta}), we 
would have found an expression that is approximately $2\pi 
r_{s}\delta(k_{z})$). Thus the predicted contrast spectrum would have too large 
an amplitude by a factor $r_{s}/l_{i}$, which would correspond to deflections 
suffered in consequence of a turbulent field pervading the entire region 
between the implosion capsule and the screen, rather than one confined to a 
small interaction region.

We Fourier-transform Equation~(\ref{eq:Correlation_4}) with respect to
both arguments, obtaining
\beq
\left\langle
\tilde{\Lambda}(\bp_\perp)\tilde{\Lambda}(\bp_\perp^\prime)^*
\right\rangle=
\frac{2\pi e^2(r_s-r_i)^2l_i}{m^2c^2v^2}
\delta^2(\bp_\perp-\bp_\perp^\prime)
E_{B}\left(\frac{r_s}{r_i}p_{\perp}\right),
\label{eq:Correlation_5}
\eeq
where we've defined the Fourier transform of the contrast fluctuation 
map,
\beq
\tilde{\Lambda}(\bp_\perp)\equiv(2\pi)^{-4}\int d^2\bx_\perp 
e^{-i\bp_\perp\cdot\bx_\perp}\Lambda(\bx_\perp).
\label{eq:Lambda_FT}
\eeq

The awkward $\delta$-function in Equation~(\ref{eq:Correlation_5}) 
may be disposed of by noting that the contrast fluctuation map 
$\Lambda(\bx_\perp)$ is available as a discretized array on a mesh 
of cells, rather than as a continuous function, and its Fourier 
transform is necessarily a Discrete Fourier Transform (DFT).  If the 
mesh on the screen is an $N\times N$ evenly-spaced array over a 
square of lateral dimension $L$, the DFT will produce an complex array
of transform values $\hat{\Lambda}(\bn)$, where the $\bn$ are 
2-dimensional mode vectors whose components are integers, 
$\bn=[n_x,n_y]$.  The $\bn$ are related to the wave vectors $\bp_\perp$ by
$\bp_\perp=\frac{2\pi}{L}\bn$.  The $\hat{\Lambda}(\bn)$ are then related to 
the $\tilde{\Lambda}(\bp_\perp)$ by averaging the latter over the cell 
in $\bp_\perp$-space corresponding to the mode $\bn$.  That is to say,
\beq
\hat{\Lambda}(\bn)=\left(\frac{2\pi}{L}\right)^{-2}\int_{\mbox{\scriptsize Cell}~\bn}
d^2\bp_\perp\,\tilde{\Lambda}(\bp_\perp),
\label{eq:cell_average}
\eeq
where the integral is over the $\bp_\perp$-space square cell of side 
$2\pi/L$ centered at $\frac{2\pi}{L}\bn$.

We therefore average the dependence of Equation~(\ref{eq:Lambda_FT}) on
$\bp_\perp$, $\bp_\perp^\prime$ over cells $\bn$, $\bn^\prime$, 
respectively.  This process duly removes the $\delta$-function, leaving
the result
\beq
\left\langle
\hat{\Lambda}(\bn)\hat{\Lambda}(\bn^\prime)^*
\right\rangle=\frac{e^2(r_s-r_i)^2l_iL^2}{2\pi m^2c^2v^2}
\delta_{\bn\bn^\prime}
E_{B}\left(\frac{r_s}{r_i}\frac{2\pi}{L}|\bn|\right)
.\label{eq:Correlation_6}
\eeq

The final result for the turbulence spectrum is then
\beq
E_{B}\left(\frac{r_s}{r_i}\frac{2\pi}{L}|\bn|\right)
=\frac{2\pi m^2c^2v^2}{e^2(r_s-r_i)^2l_iL^2}
\left\langle
\hat{\Lambda}(\bn)\hat{\Lambda}(\bn)^*
\right\rangle.
\label{eq:Spectrum}
\eeq
In a practical computation, the ensemble average in 
Equation~(\ref{eq:Spectrum}) may be replaced by a rotational average 
over cells in $\bn$-space, binning $|\bn|$ to suit one's convenience.

\begin{figure*}[t]
\begin{center}
\includegraphics[width=7.0in, viewport=175 65 1660 820, clip=]{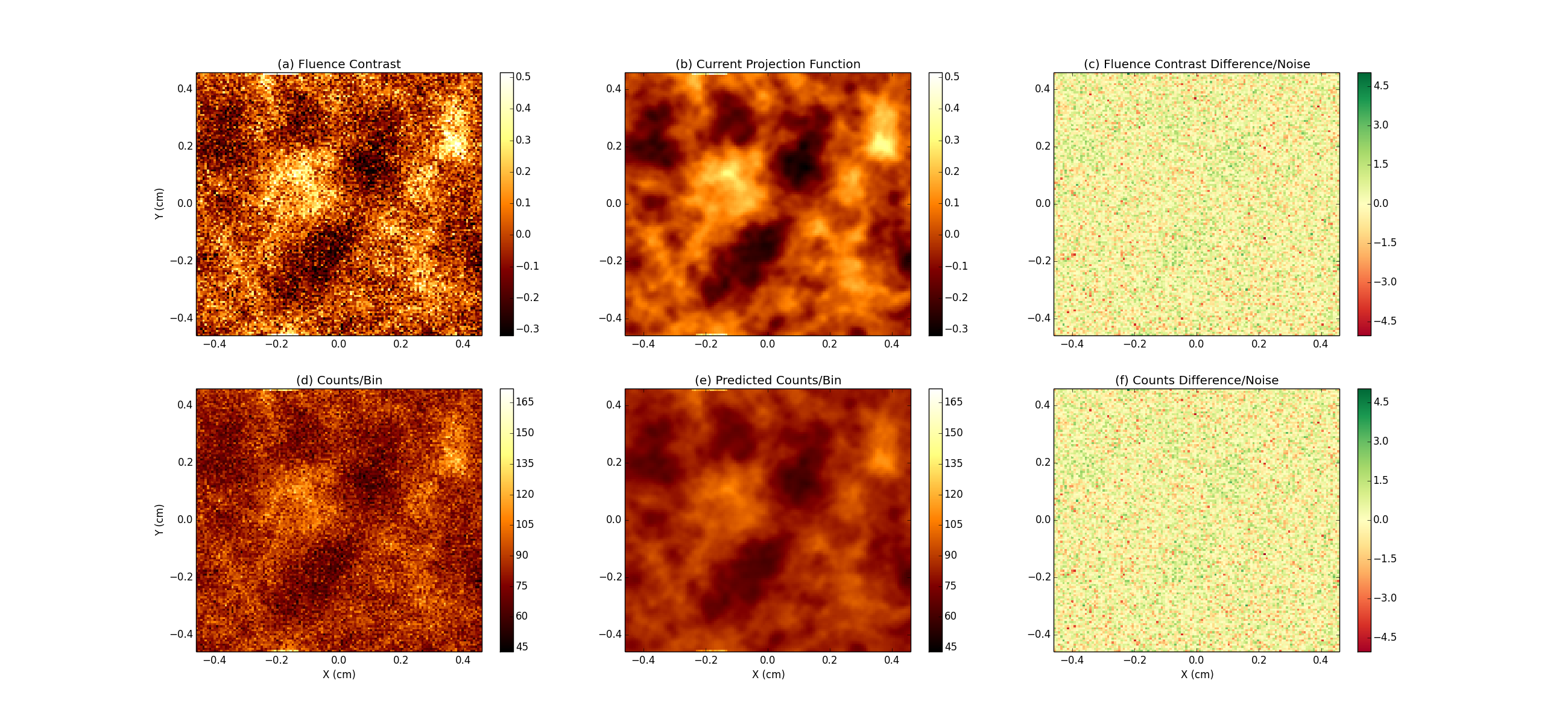}
\caption{\label{figure:GPRadiograph}Radiographic results of simulation with 
Gaussian random field. The images are of a square of data entirely contained 
within the image of the interaction region.}
\end{center}
\end{figure*}

\begin{figure*}[t]
\begin{center}
\includegraphics[width=6.8in, viewport=75 0 975 431, clip=]{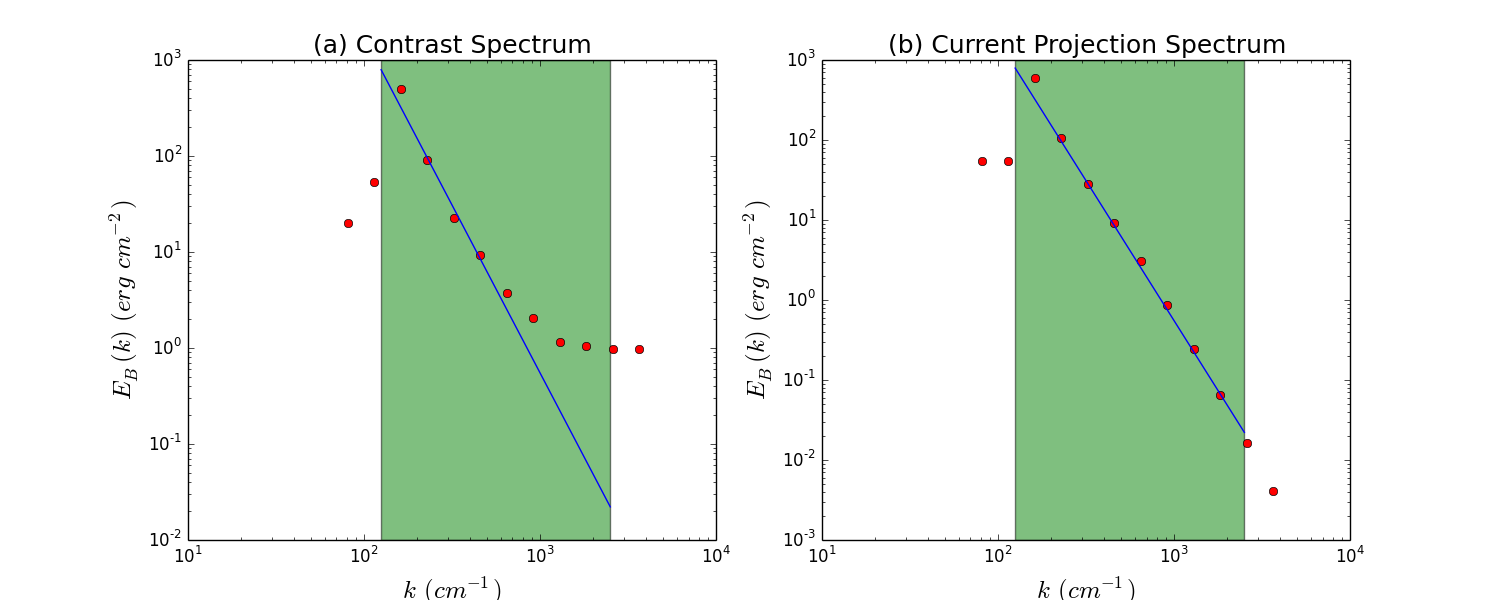}
\caption{\label{figure:spectrum} Recovered spectra using Equation~(
\ref{eq:Spectrum}). In both panels the highlighted region is the 
range between the spectral cutoffs $k_1$ and $k_2$, while the blue 
line represents the input spectrum that was used to generate the 
field according to Equation~(\ref{eq:genspec}).  (a) Spectrum 
obtained using the binned contrast data.  (b)
Spectrum obtained repeating the procedure, but using the current 
projection function $\chi$ (which predicts the contrast) instead.  
The high-frequency differences between the two spectra are explained 
by Poisson noise.}
\end{center}
\end{figure*}

It is a noteworthy feature of Equation~(\ref{eq:Spectrum}) that the factor 
$\frac{r_{s}}{r_{i}}p_\perp$ in the argument of $E_B$ expresses the 
expected stretching of wavelengths by the divergence 
of the proton rays on their way from the interaction region to the 
screen.

Note also that according to Equation~(\ref{eq:Spectrum}), we again require an 
estimate of the longitudinal extent of the interaction region $l_i$ in order to 
obtain the amplitude of the spectrum $E_B$.  The discussion of the availability of such an 
estimate of $l_i$ was given at the end of \S\ref{sec:reconstruction}.

\subsection{Numerical Verification with a Gaussian 
Field}\label{subsec:GP_Verification}

We verify Equation~(\ref{eq:Spectrum}) by means of a simulation in which 
protons are fired at a domain containing a Gaussian random magnetic field of 
known spectrum.  Such a field does not really represent a realistic magnetic 
turbulence distribution, of course, since turbulence is well-known to be 
non-Gaussian (see \textcite{davidson2004turbulence}, pp. 99-103, for example). This 
is not an issue here, since we merely wish to confirm that we can successfully 
recover the two-point spectrum of a homogeneous-isotropic distribution, 
irrespective of what the other moments of the distribution may be, so we are 
free to choose a Gaussian random field on grounds of simplicity.

The optical set-up is the same as in the previous verification tests: 
$r_i=10$~cm, $r_s=100$~cm, $l_i=0.1$~cm.  Protons energies are monochromatic, 
at 14.7~MeV, and the protons are emitted from a point source at the origin.

The simulated spectrum is a truncated power-law with the following form:
\beq
E_B(k)=
\left\{
\begin{array}{c@{\quad:\quad}l}
\frac{\left<B^2\right>}{8\pi}\frac{\alpha+1}{k_2^{\alpha+1}-k_1^{\alpha+1}}k^\alpha
& k_1<k<k_2\\
0&\mbox{otherwise}.
\end{array}
\right.\label{eq:genspec}
\eeq
This form allows us to specify the normalization in terms of the mean magnetic 
energy $B^2/8\pi$, according to Equation~(\ref{eq:Espectrum}).  We parametrize 
the spectral cutoffs in terms of discrete modes $n_1$ and $n_2$, with 
$k_1=\frac{2\pi}{l_i}n_1$, $k_2=\frac{2\pi}{l_i}n_2$.  In our simulations, we 
chose $<B^2>^{1/2}=10^3$~G, $\alpha=-3.5$, $n_1=2$, $n_2=40$.  These 
parameters, together with the optical system parameters $r_s$, $r_i$ above, 
result in expected contrast fluctuations at the screen of about 0.1.

We generate the field in a $128\times128\time128$ mesh of cells in a cubic 
domain.  We start out in Fourier space, randomly drawing the (complex) 
components of the Fourier-transformed vector potential $\ba(\bk)$ in each 
$\bk$-space cell indepenently from a zero-mean normal distribution. The 
variance of the distribution is chosen to yield the desired spectrum $E_B$.  
The reality constraint $\ba(\bk)^*=\ba(-\bk)$ is enforced by drawing samples 
only in the hemisphere $k_x>0$, and copying the complex conjugate of the field 
in this hemisphere to the reflected point in the opposite hemisphere. The 
resulting vector field is then Fourier-transformed to the real domain to obtain 
$\bA(\bx)$, and the discrete curl is taken to yield the final magnetic field, 
$\bB=\nabla\btimes\bA$.

We simulate firing 5 million protons through this set-up, in a cone of radius 
0.1 cm at $r=r_i$.  The radiographic results are displayed in 
Figure~\ref{figure:GPRadiograph}, which shows a square region of data entirely 
contained within the image of the interaction region.

To obtain the spectrum, we proceed as indicated in the discussion following 
Equation~(\ref{eq:Spectrum}).  We bin the region imaged in 
Figure~\ref{figure:GPRadiograph} into a $128\times 128$ array of cells, 
computing the contrast map in each cell, and perform an FFT on the result, to 
estimate $\hat\Lambda(\bn)$.  We bin the radial wave number $n=|\bn|$ into equally-spaced
logarithmic bins, and average the contributions in each bin to perform a 
circular average.  Then, using Equation~(\ref{eq:Spectrum}), we estimate $E_B$.

The result of this computation is displayed in Figure~(\ref{figure:spectrum}). 
The left panel of this figure shows the result of the analysis just described. 
The right panel shows the result of repeating the analysis on the current 
projection function $\chi$, which as we have seen, predicts the fluence 
contrast.  It is clear that at low frequency, the analysis returns the correct 
spectrum -- both the low-frequency cutoff $k_1$ and the spectral slope are 
correctly recovered.  At high frequency, we see a leveling-off of the spectrum 
in the contrast data, that has no counterpart in the spectrum inferred from 
$\chi$.  Since the difference between these two is basically Poisson noise, we 
see that it is the white-noise spectrum from Poisson statistics that supervenes 
to level off the decline of the magnetic spectrum.

This observation points to a practical necessity in implementing this 
algorithm: it is necessary to optimize the tradeoff between spectral resolution 
and Poisson noise, by choosing the size of the image binning mesh carefully. 
Too high a resolution can lead to low counts-per-bin, and therefore excessive 
noise, while too low a resolution leads to the inability to probe high 
frequencies.  This tradeoff must be resolved empirically in the analysis of 
radiographs, on a case-by-case basis.  These considerations are 
analogous to the ones described in \S\ref{subsec:twoblobs}.

Another high-frequency effect that is to be expected in real data, but that we 
do not simulate here, is that the finite size of the implosion capsule must 
necessarily produce a high-frequency cutoff in the measured spectrum.  The 
reason is that a finite capsule size $s$ blurs the image of every point in the 
interaction region by an angular blurring $s/r_i$, and hence by a linear 
blurring length $s(r_s-r_i)/r_i$ in the image.  This corresponds to a length 
scale in the interaction region of $\lambda_s=s(r_s-r_i)/r_i\times 
r_i/r_s=s(r_s-r_i)/r_s$. This is the limit to which we may expect to resolve 
any structure in the spectrum, which should therefore be expected to contain a 
high-frequency cutoff at wavenumber $k_s=2\pi/\lambda_s$.  Naturally, for this 
cutoff to be visible, it must occur at wavenumbers $k$ below the advent of the 
white noise plateau discussed above.

\section{On The Nonlinear Theory of Proton Radiography}\label{sec:Nonlinear}

An important gap that remains in the analysis of 
proton radiographs is the ability to reconstruct the transverse 
magnetic field in the nonlinear regime in a manner analogous to the 
linear reconstruction discussed above.  From an experimental point 
of view, this is an important gap to address, since radiographs of 
laser plasmas can show evidence of nonlinear contrasts (see, for 
example, Figure 6 of \textcite{Park2015}).  In these cases, the 
linear theory can provide at best crude and approximate estimates of 
field strength and morphology.

While we do not furnish here a finished theory -- complete with 
numerical verification -- of nonlinear radiography, we believe that we 
know enough about the features of such a theory (as distinct from the 
caustic-based interpretive theory of K2012) to be worth setting out in
this article.

The starting place is the nonlinear equation of intensity at the 
screen.  This equation, which replaces the linear regime 
Equation~(\ref{eq:Proton_Cons2}), is given by Equation~(6) of K2012, 
which we write here as 
\bea
\Psi\left(\bx^{(0)}+\bw\right)
\times\det\left|
\frac{\partial (\bx^{(0)}+\bw)}{\partial\bx^{(0)}}
\right|=\psi_0.\label{eq:Intensity}
\eea
Here, as before, $\Psi$ is the measured fluence on the radiograph, $\psi_0$ is
the fluence that would be measured in the absence of any deflection, 
$\bx^{(0)}$ is the location on the screen where a proton would land in the absence of 
deflection, and $\bw$ is the lateral deflection.

Now, even in the nonlinear regime, it remains the case that 
$\bw=-\nabla\phi$, as was discussed in \S\ref{sec:reconstruction} -- this is 
essentially a consequence of the irrotationality of the magnetic field.  If we 
define the function
\bea
\zeta\equiv \frac{1}{2}|\bx^{(0)}|^2 - \phi,
\label{eq:potential}
\eea
then we may write Equation~(\ref{eq:Intensity}) in the following form:
\bea
\Psi\left(\nabla\zeta\right)
\times\det\left|
D^2\zeta
\right|=\psi_0,\label{eq:MongeAmpere}
\eea
where $D^2\zeta$ is the matrix of second derivatives of $\zeta$, that is 
$[D^2\zeta]_{ij}=\partial^2\zeta/\partial x^{(0)}_i\partial x^{(0)}_j$.

As intractable as this equation appears at first glance, it turns out to be an 
example of a very famous class of equations:  it is the \textit{Monge-Amp\`ere
equation}, first discussed by the French mathematicians Gaspard Monge (in 1784) 
and Andr\'e-Marie Amp\`ere (in 1820).  It is an equation of central importance in 
the subject of Monge-Kantorovich Optimal Mass 
Transportation\cite{evans1997}, and it crops up in fields as diverse as 
differential geometry\cite{caffarelli1999}, meteorology\cite{douglas1999}, 
cosmology\cite{frisch2002}, medical imaging\cite{haker2004}, economics\cite{blanchet2014}, and many others.  For 
this reason, the properties of this equation have been studied in depth.

The equation arises in its archetypal form in optimal transportation 
problems, wherein a fixed given distribution of mass in some space is to 
be mapped to a second fixed distribution of mass, by means of a mapping on 
the space that minimizes some cost function.  In symbols (if not in 
formal mathematics) the initial distribution $\mu_0$, a measure on a 
space $\Omega$, is to be mapped to a new, given distribution $\mu_1$ 
on $\Omega$ by means of a ``transport plan'', a diffeomorphism 
$\by:\Omega\rightarrow\Omega$, so that $\mu_1$ and $\mu_2$ are 
related by the jacobian of $\by(\bx)$.  Obviously, there are many 
possible transportation plans that can effect the required mapping, 
but the one of interest is the one that minimizes a certain cost 
function.

Suppose that the distributions $\mu_0$, $\mu_1$ are described by 
densities $P_1(\bx)d\bx$, $P_2(\bx)d\bx$, respectively.  $P_1$ and 
$P_2$ are related by
\bea
P_2(\by)\det\left|\frac{\partial\by}{\partial\bx}\right|=P_1(\bx).
\label{eq:volpreserve}
\eea
If the cost function to be minimized is the 
\textit{Kantorovich-Wasserstein $L2$ distance}
\bea
d_2(P_1,P_2,\by)\equiv\int_\Omega d\bx\, P_1(\bx) (\by(\bx)-\bx)^2,
\label{eq:KW_distance}
\eea
then it has been shown\cite{brenier1991,gangbo1996} that the 
cost-minimizing diffeomorphism $\by(\bx)$ is the gradient of a convex 
function $\zeta$, that is,
\bea
\by(\bx)=\nabla\zeta(\bx).
\label{eq:Is_a_gradient}
\eea

Inserting Equation~(\ref{eq:Is_a_gradient}) in Equation~(\ref
{eq:volpreserve}), it then follows that $\zeta(\bx)$ is the solution 
of a Monge-Amp\`ere equation.  It has been further shown that this 
solution is unique.

There is therefore a curious connection between the nonlinear 
radiography reconstruction problem and the optimal transportation 
problem.  Whereas in the latter, it is the case that the $L2$-cost 
minimizing optimal diffeomorphism is the gradient of the unique 
solution of the Monge-Ampere equation, in the radiography problem, 
we know that the mapping between the no-deflection fluence 
distribution and the observed distribution is necessarily the 
gradient of some function (by the irrotationality of the magnetic 
field), and that therefore the unique solution of the Monge-Amp\`ere 
equation (\ref{eq:MongeAmpere}) is the optimal -- by the $L2$ cost 
metric -- transport plan between the uniform fluence density $\psi_0$
and the observed density $\Psi$.  

In other words, in our current notation, given the cost 
function
\bea
d_2(\Psi,\psi_0,\bw)=\psi_0\int d^2\bx\,\,|\bw(\bx)|^2,
\label{eq:Cost}
\eea
optimal transportation theory tells us that minimizing Equation~(\ref
{eq:Cost}) with respect to the set of deflections $\bw(\bx)$ that transform 
$\psi_0$ to $\Psi(\bx)$ leads to an irrotational vector field 
$\bw(\bx)=-\nabla\phi(\bx)$, and 
hence to a solution of Equation~(\ref{eq:MongeAmpere}).  

This connection creates an opportunity.  There are a number of 
numerical algorithms in the literature 
\cite{benamou2010,haker2004,rachev1998,cullen1984extended} that have been 
successfully used to solve the optimal transportation problem with 
$L2$ cost in various contexts.  The successful application of any 
such algorithm to the radiographic inversion problem would produce 
the lateral deviations $\bw$, and hence, by Equation~(\ref 
{eq:Lateral_Deviation2}), the average transverse magnetic field.

This discussion of the nonlinear inversion problem is necessarily 
schematic, and is clearly more a description of a possible research 
path than actual research in its own right.  The connection between 
nonlinear radiography and optimal transportation theory seems to us 
to be a valuable result, worth mentioning in print, and we hope it 
will be helpful in stimulating further research on the topic.

\section{Discussion}\label{sec:Discussion}

We have shown in this work several new and interesting results in the theory of 
proton radiography.  First, we have shown that in the linear 
contrast fluctuation regime, proton radiographic images of magnetized 
(non-electrified) plasmas are simply \textit{projective images of MHD current}. 
This fact, which was not previously recognized, lends considerably more 
diagnostic power to the analysis of such radiographs, which may be interpreted 
directly in terms of the currents that they image.  This is a much more direct 
and less cumbersome analysis than that proposed in K2012\cite{kugland2012a}, 
wherein radiographs are interpreted using stacks of blob-like field 
configurations chosen to mimic structures observed in the image.

Secondly, we have shown that it is possible to reconstruct the line-integrated 
transverse magnetic field, by solving a steady-state diffusion equation with a 
source and an inhomogenous diffusion coefficient that are both functions of the 
the proton contrast image on the radiograph.  Given an independent estimate of 
the interaction region size $l_i$, this translates directly into estimates of 
the transverse magnetic field, averaged over proton trajectories.

Thirdly, we have shown that proton-radiographic images of isotropic-homogeneous 
turbulent magnetic fields can be straightforwardly analyzed to infer the 
field spectrum, which is obtainable from the Fourier transform of the two-point 
autocorrelation function of the image.  The average magnetic field energy 
density is also easily obtainable from this kind of analysis.

Fourth, and perhaps most importantly, we have propounded a view of proton 
radiograph analysis that is rather different from what has become a standard 
view, to wit that structures on radiographs are explainable by caustic 
formation, and that identification of such structures and their analysis in 
terms of caustics is necessary to estimate magnetic structure from images.  
That view became prevalent, we believe, because the direct relation between 
radiographic structure and MHD current structure was not yet understood, and it 
was felt that the explanation of sharply-defined structures on radiographs 
could not be carried out merely in terms of presumably-smooth magnetic fields.  
For this reason, appeal was made to caustic properties of optical systems, to 
describe the observed sharply-defined structures.

In the view described in this work, while caustics are undoubtedly a possible 
feature of proton radiographic images, it is not generally necessary to focus 
on caustic structure to interpret such images.  As we have shown, sharp current 
features such as filaments and sheets automatically result in analogous sharp 
structures on the radiograph, whose presence may therefore not be the result of caustic 
structure.  At least in the linear contrast regime the transverse 
magnetic field is directly recoverable from the image (as is the energy 
spectrum, in the case of homogenous-isotropic turbulence). 
Furthermore, we have suggested in \S\ref{sec:Nonlinear} an approach 
that may allow field reconstruction even in the nonlinear regime.
Therefore, the approach of 
modeling various component blobs and shocks and comparing their radiographic 
signatures to observed structures seems less compelling.  In any event, from an 
aesthetic point of view, it seems to us much more satisfactory to be able to 
interpret radiographic images not as entangled products of the plasma structure 
and the focusing properties of the proton optical system, but rather as the 
direct result of the plasma structure alone.

\section{Acknowledgements}
This work was supported in part at the University of Chicago
by the U.S. Department of Energy (DOE) under contract B523820 to
the NNSA ASC/Alliances Center for Astrophysical Thermonuclear
Flashes; the Office of Advanced Scientific Computing Research,
Office of Science, U.S. DOE, under contract DE-AC02-06CH11357;
the U.S. DOE NNSA ASC through the Argonne Institute for
Computing in Science under field work proposal 57789; and the
U.S. National Science Foundation under grant PHY-0903997.
Carlo Graziani would like to thank Fausto Cattaneo for very useful 
advice on numerical matters.

\section*{References}

\bibliography{radiography}

\end{document}